# VistaScenario: Interaction Scenario Engineering for Vehicles with Intelligent Systems for Transport Automation

Cheng Chang, Jiawei Zhang, Jingwei Ge, Zuo Zhang, Junqing Wei,
Li Li, *Fellow, IEEE*, and Fei-Yue Wang, *Fellow, IEEE*

*Abstract*— **Intelligent vehicles and autonomous driving systems rely on scenario engineering for intelligence and index (I&I), calibration and certification (C&C), and verification and validation (V&V). To extract and index scenarios, various vehicle interactions are worthy of much attention, and deserve refined descriptions and labels. However, existing methods cannot cope well with the problem of scenario classification and labeling with vehicle interactions as the core. In this paper, we propose *VistaScenario* framework to conduct interaction scenario engineering for vehicles with intelligent systems for transport automation. Based on the summarized basic types of vehicle interactions, we slice scenario data stream into a series of segments via spatiotemporal scenario evolution tree. We also propose the scenario metric Graph-DTW based on Graph Computation Tree and Dynamic Time Warping to conduct refined scenario comparison and labeling. The extreme interaction scenarios and corner cases can be efficiently filtered and extracted. Moreover, with naturalistic scenario datasets, testing examples on trajectory prediction model demonstrate the effectiveness and advantages of our framework. *VistaScenario* can provide solid support for the usage and indexing of scenario data, further promote the development of intelligent vehicles and transport automation.**

## I. INTRODUCTION

W ITH the development of vehicles with intelligent systems for transport automation (Vista), autonomous driving has been applied in various domains [1]-[3]. During the intelligence and index (I&I), calibration and certification (C&C), and verification and validation (V&V) [4] process of autonomous systems and functions, scenario engineering plays a vital role on the assurance of the "6S" objectives (Safety, Security, Sustainability, Sensitivity, Smartness, and Services) [5]-[7].

Driving scenarios [8]-[10] are generally defined as the comprehensive reflection of the environment and driving behaviors within a certain temporal and spatial scope. Originated from scenarios, scenario data are the explicit integration of road environment data and traffic participants data. Within the scenario information, vehicle interactions are worthy of much attention, which can reduce the driving risk and improve the traffic flow [11]-[12].

To make full use of scenario data, the descriptions and associated interaction labels are intuitively important. Many autonomous driving modules, such as trajectory prediction [13]-[15], motion planning [16]-[18], and intelligence testing [19]-[20], all need the feed of scenario data. On the one hand, the modules need to accurately extract complex patterns and interactions of traffic participants. To train and optimize the modules, it is essential to obtain appropriate descriptions and labels as supervision. On the other hand, the modules should adapt to different types of scenarios and environments. We need to balance scenario samples to maintain generalization, and sometimes select many extreme and challenging scenarios with complex interactions to verify the designed algorithm. Therefore, clear scenario classification levels and efficient scenario comparison methods that specially consider interactions are needed to help manage and categorize large amounts of scenario data.

To support the I&I, C&C, and V&V goals of scenario engineering, we need specially consider three crucial questions. First, *How to properly consider vehicle interactions in scenario classification?* Vehicle interactions have complex mechanisms, and vary with driving strategies. However, considering numerous parameters of the interactions will cause high computational complexity and cannot achieve complete classification [21]-[22]. To simplify the issue and accord with human driver's experience, when vehicles interact with others, they need to check the potential conflict area and collisions, and communicate to determine the right-of-way [23] to maintain an appropriate balance between safety and efficiency. Based on the simplified interaction process, we can consider interactions mainly according to the conflict relations of vehicle motion flows. Some basic interaction types should be summarized to slice the complex interaction segments. In this way, the quantity of interaction types, interactive vehicles, and corresponding scenario samples can be enumerated thoroughly.

This work was supported in part by the National Key Research and Development Program of China under Grant 2023YFB2504400. (Corresponding author: *Li Li*)

Cheng Chang, Jiawei Zhang, Jingwei Ge, and Zuo Zhang are with the Department of Automation, Tsinghua University, Beijing 100084, China.

Junqing Wei is with DiDi Autonomous Driving Company, Beijing 100094, China.

Li Li is with the Department of Automation, BNRist, Tsinghua University, Beijing 100084, China (e-mail: li-li@tsinghua.edu.cn).

Fei-Yue Wang is with the State Key Laboratory for Management and Control of Complex Systems, Institute of Automation, Chinese Academy of Sciences, Beijing 100190, China.

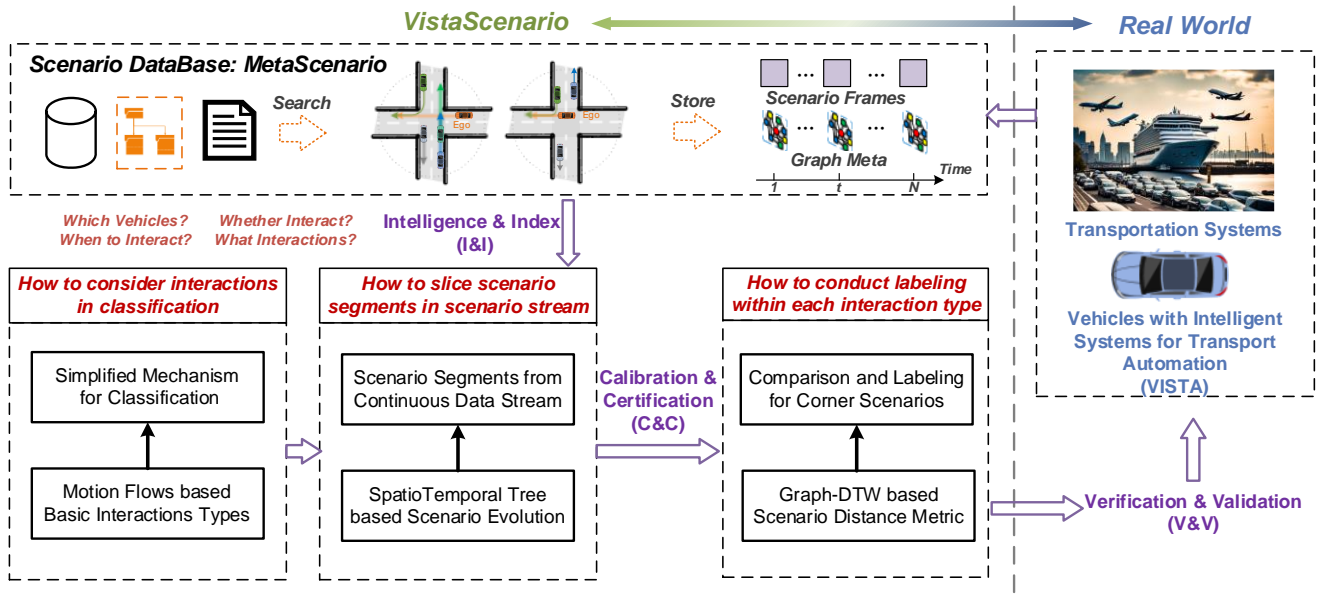

Fig. 1. The illustration of VistaScenario framework.

Second, *How to slice scenario segments in scenario data stream?* With the movements of vehicles, scenario data stream is constantly generated. Some researches regard the scenarios as a whole set and sample typical data to constitute a new sampled dataset [24]-[26]. The verification and evaluation of algorithms are conducted on the sampled set instead of the whole set to accelerate the process. However, the methods imply homogeneous driving behaviors and lack the distinctions for vehicle maneuvers and interactions. When conducting refined scenario analysis, the scenario data stream should be sliced into series of segments with different types of vehicle interactions. It is comprehensible to describe driving scenarios with high level behaviors and interactions, which greatly reduces the number of scenario variables. Scenario segments with detailed annotations of interactions are also more practical to be applied to autonomous driving modules. In previous researches [27]-[28], tree search-based methods have shown the effectiveness for processing stream data, which can represent spatiotemporal vehicle interactions and movement schedules.

Specifically, the above two questions, i.e., interaction-based scenario data slicing and classification solve the following "4W" questions: 1) *Which Vehicles?* We need to search which vehicles are moving in the same spatial area as ego vehicle. 2) *Whether Interact?* It needs to be judged whether the searched vehicles have conflicts with the ego at the moment or in the near future based on the motion flow. 3) *When to Interact?* If the vehicle interaction is detected, we should further check whether the vehicle is in the interaction process at current frame, and extract the interaction beginning and ending time. 4) *What Interaction?* It is necessary to clarify what the interaction types between surrounding vehicles and ego vehicle, and whether the continuous frames can be merged into one scenario segment which requires relatively consistent interaction types. The four aspects of content constitute the attributes of scenario segments.

Third, *How to conduct refined comparison within each interaction type?* Even if the scenario segments are of the same interaction type, the differences also exist. It is necessary to further conduct refined comparison and labeling under certain interested criteria. The most typical and attractive one is driving risk, which is close related to vehicle interactions [4], [29]. In the training and testing stage, we can prioritize the extreme scenario samples for better performance and generalization. However, scenario segments with complete interaction process usually correspond to different time lengths, which makes it inappropriate to make simple alignment comparisons. Although Dynamic Time Warping (DTW) [30]-[31] method, which supports multi-time correspondence computation, can properly handle the question, existing DTW-based methods usually only consider the composition of some vehicle states. For scenario segments contain different types of driving environment and different numbers of vehicles, we need a more general and suitable structure to efficiently represent vehicle interactions. Within many data structures, graph formats can represent interaction scenarios semantically and flexibly [8], [32]-[33].

To handle the above three questions, in this paper, we design *VistaScenario* framework to conduct the interaction scenario engineering for the I&I, C&C, V&V of vehicles with intelligent systems for transport automation (Vista). The *VistaScenario* framework is an expansion of our previous work *MetaScenario* [8], which focuses more on vehicle interactions. The illustration of processing flow is shown in Fig. 1. The key contributions are as follows:

1) To utilize the interaction information, we provide the scenario classification hierarchy related to vehicle interactions. We achieve complete and efficient classification via simplified interaction mechanism. The basic types of vehicle interactions are extracted and summarized based on the vehicle motion flows. The approach allows us to synthetically consider the interaction flow of vehicles and the road/lane topology. We also analyze the interaction statistics on *HighD* and *INTERACTION* naturalistic scenario datasets.

2) To efficiently slice the scenario segments, we design the

spatiotemporal evolution tree. The segments are preliminarily classified based on interaction types, which contain complete and consistent decision-making process. Relative to the ego, the interactive vehicles and interaction durations are effectively extracted. Non-interactive vehicles are filtered out, which facilitates the compression of scenario space. The methods can address the "4W" interaction questions mentioned above.

3) To conduct scenario comparison and labeling within each interaction type, we propose the Graph-based Dynamic Time Warping (Graph-DTW) metric. We can obtain the distance of each scenario pair at typical aspects, such as driving risks. Compared with other metrics, Graph-DTW can efficiently calibrate and certificate more extreme scenarios and corner cases. The constructed library can further facilitate the C&C and V&V of transport systems and Vista in the real world. Testing examples on prediction model show the advantages of our proposed method. We also provide potential applications for Retrieval Augmented Generation (RAG) in Large Language Model (LLM).

The rest of this paper are arranged as follows. *Section II* reviews related works. *Section III* summarizes the basic types of vehicle interactions. *Section IV* introduces the segment and classification of scenarios. *Section V* introduces the refined scenario distance metric. *Section VI* presents the scenario segments statistics and classification results of typical naturalistic datasets. The advantages and applications of the metric are explained. Finally, *Section VII* concludes the paper.

## II. RELATED WORKS

In recent years, there have been various methods and algorithms proposed for driving scenario classification and labeling. The methods can be broadly categorized into two types, road environment classification and vehicle behavior classification.

### A. Road Environment Classification

Road environments are the foundation of scenarios, and researchers mainly classify road environments according to lane topology and road geometry. One typical tool OpenDRIVE [34] provides the attributes and annotations that can be used for classification. The attributes can describe road types (urban, highway, rural), road geometry (intersections, roundabouts), and polynomial-based lane characteristics (number of lanes, lane width). Another typical tool LaneLet [35] is also tagged with various attributes and traffic elements, which include left-right boundary-based lanes, traffic signals, stop lines, crosswalks, etc. However, the road environment mainly contains the static scenario information, thus the classification levels cannot cover the dynamic evolutions of vehicle behaviors and interactions.

### B. Vehicle Behavior Classification

The behaviors and interactions of vehicles are important dynamic scenario components. Vehicle behavior classification is further categorized into the types of behavior characteristics, single vehicle maneuver, and multi-vehicle interaction classification.

1) **Behavior Characteristics Classification:** Considering various driving behavior characteristics, researchers tend to use the parameters strongly related to the criteria to classify the scenarios. For example, the driving safety metrics, Time-To-Collision (TTC), Time Headway (THW), critical states are used to estimate and extract the risky driving scenarios [36]-[38]. The traffic flow metrics, vehicle speed and density, are used to classify the scenarios under dense and sparse vehicle flow [39]-[40]. However, the parameters are always fixed and may lack accuracy and flexibility in some types of driving environments [41]. The methods only address simple vehicle movements and cannot well handle vehicle interactions. Moreover, they mainly help obtain the extreme scenes rather than complete scenarios, so the question "slicing the scenario segment" cannot be clearly solved.

2) **Single Vehicle Maneuver Classification**: Compared to critical driving characteristics, many researchers focus more on concrete maneuvers and behaviors. For single vehicles, several rules are typically defined by experts or engineers, and can be used to make real-time decisions and classifications. The rules usually utilize ontology concepts and relationships. For instance, if the environment is intersection and traffic light is red, then the scenario will be classified as "vehicle stops at intersection". Similar semantic rules are applied to constitute some scenario libraries. [42]-[43] construct semantic map and match the vehicle trajectory points with the map to obtain corresponding maneuvers, such as Right Turn or Cross Junction. [44] proposes spatiotemporal ontology for describing vehicle movements into directional and longitudinal descriptions, and then classifies the maneuvers. Although the methods have strong interpretability for different scenarios, the limited flexibility will struggle with handling complex or ambiguous scenarios that fall outside the predefined rule set.

3) **Multi-Vehicle Interaction Classification**: Scenarios usually involve complex multi-vehicle interactions and need further information for classification. According to whether the methods rely on interaction and behavior labels, they can be mainly categorized into supervised methods and unsupervised methods.

*Supervised Methods*: Deep neural network-based methods such as Convolutional Neural Network (CNN) [45], Recurrent Neural Network (RNN) [46], and Graph Neural Network (GNN) [47] are used to encode the interactions among vehicles, then distinguish and classify the behaviors and interactions. However, the supervised data-driven methods rely on necessary annotation and labels for training. For most collected scenario datasets, the lack of enough labels and attributes itself creates a bottleneck in the implementation and generalization of the methods. The rise of foundation models, such as GPT [2], [48] and Sora [49]-[51], which are pretrained on massive amount of data, gives the potential to recognize the vehicle behaviors automatically. The models have strong scenario reasoning and understanding ability to handle complex scenarios. Besides, the models can process multi-modal data, such as text, images and video, which facilitate to analyze the diverse information required for vehicle interactions. However, the outputs of large models are sometimes uncertain and unreasonable. It is also still expensive to train and call the large models in current stage.

*Unsupervised Methods*: In this type of methods, the scenario information is summarized into composed vectors or special structures for analysis without supervised label information. As typical examples, [52] uses eight-vehicle slot distance matrix around ego to calculate scenario distance, [53] selects features composed of 10 types of road layer and object layer features to represent scenarios. Unsupervised mining algorithms such as Clustering [52], Random Forests [53], Autoencoders [30] are used to compare and classify the driving scenario. However, different scenarios contain different number of interactive vehicles. The road network topology is also complex, and may not be regular straight function. Thus, the fixed encodings for vehicles and road/lanes are not so appropriate in above cases, the third question related to vehicle and road interactions cannot be well handled. To flexibly and directly represent and compare scenario segments, graph structures are more suitable than others, which can clearly express scenario elements and the relations among them [8], [32], [33]. For graph comparisons, one typical metric is based on graph embedding. Researchers embed the graph data into low dimensional space, and then apply the specified function (Radial Basis Function [54], Sigmoid [55], etc.) to distinguish them. However, the methods lose much structure information. Another typical metric is based on graph kernel, such as random walk [56] and subgraph isomorphism [57]. However, the expressive power of the methods is limited for the attribute graphs under driving scenario related researches. Existing graph metrics cannot support the third question, i.e. refined scenario comparisons. A more suitable graph metric for driving scenarios labeling should be properly designed.

In addition, the above methods usually predefine the spatial or temporal scope to extract the scenario segments, thus the contained vehicle interaction processes will not be complete or independent. The first and second questions, i.e., slicing and enumerating scenario segments that correspond to concrete interaction types, cannot be well addressed.

In this paper, a hierarchical scenario classification and labeling framework is proposed. First, we use scenario spatiotemporal evolution tree to slice and classify scenarios based on basic interaction types and vehicle motion flows. The mechanism can couple vehicle interaction flows with road environment, which addresses the first and second questions. Second, an unsupervised scenario distance metric Graph-DTW that can dynamically represent vehicle-to-vehicle and vehicle-to-road interactions is applied to conduct refined comparison and labeling, which addresses the third question.

## III. THE BASIC TYPES OF VEHICLE INTERACTIONS

To classify the scenarios, according to the procedure of Fig.1, we summarize the types of possible interaction flows during the movement of vehicles.

As shown in Fig. 2, first, we classify the interaction areas into two types: dynamic interaction area and static interaction area.

Dynamic interaction areas are the areas where conflicts and collisions may occur due to the various maneuvers of vehicles. The areas are usually localized, and their positions and sizes will vary with the goal and movement of vehicles dynamically. The right-of-way will also be dynamically determined with the scenario evolution. When vehicles complete the maneuvers and interaction process, the dynamic conflict areas also disappear.

Static interaction areas are the relatively fixed areas where a number of vehicles interact with others and adjust the motions. Static interaction areas are usually the critical bottleneck areas caused by the road geometry changes (e.g. ramps and intersections) or the occupancy of road sources [58]. Due to the complexity of environment, researchers also tend to make complicated cooperative driving plans and deliver the plans to vehicles in the interaction areas [59]-[60].

The basic vehicle interaction flow in dynamic interaction areas can be categorized into following types:

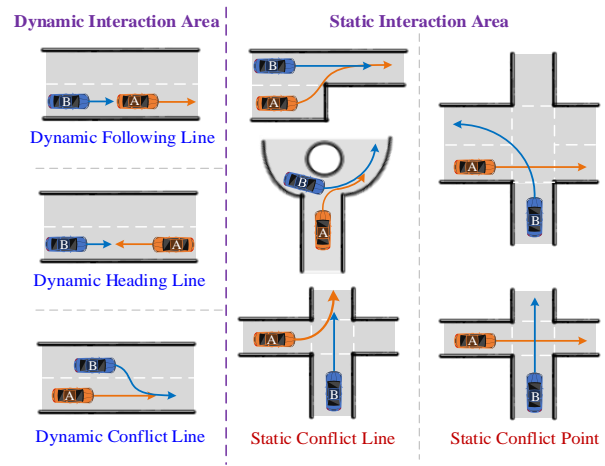

Fig. 2. The basic types of vehicle interactions.

*1) Dynamic Following Line.* The common following flow of vehicles. The zones corresponding to the car-following spacing are in the driving direction of the following vehicle. So the interaction flow is from the following vehicle to the preceding vehicle.

*2) Dynamic Heading Line.* The head-on flow of vehicles. The driving directions of two vehicles are opposite, and the interaction flow is mutual. The cases are mainly caused by vehicles traveling the wrong way down the road, or unclear lanes demarcation.

*3) Dynamic Conflict Line.* It generally refers to the flow of seizing other vehicles' right-of-way during vehicle movements, and the most common behavior is lane changing. In the decision-making process of dynamic conflict line interaction, the ego vehicle needs to interact with the vehicles of current lane and adjacent lane, and consider when and where to conduct the task.

The basic vehicle interaction flow in static interaction areas can be categorized into following types:

*1) Static Conflict Line.* It refers to the conflict lines of vehicles in static interaction areas, and the common behavior is merging, i.e., moving from different lanes to the same lane under road constraints. In the decision-making process from entering into the control area to the completion of the task, the ego vehicle interacts with other vehicles that have the same conflict area goal, and mainly considers when to pass.

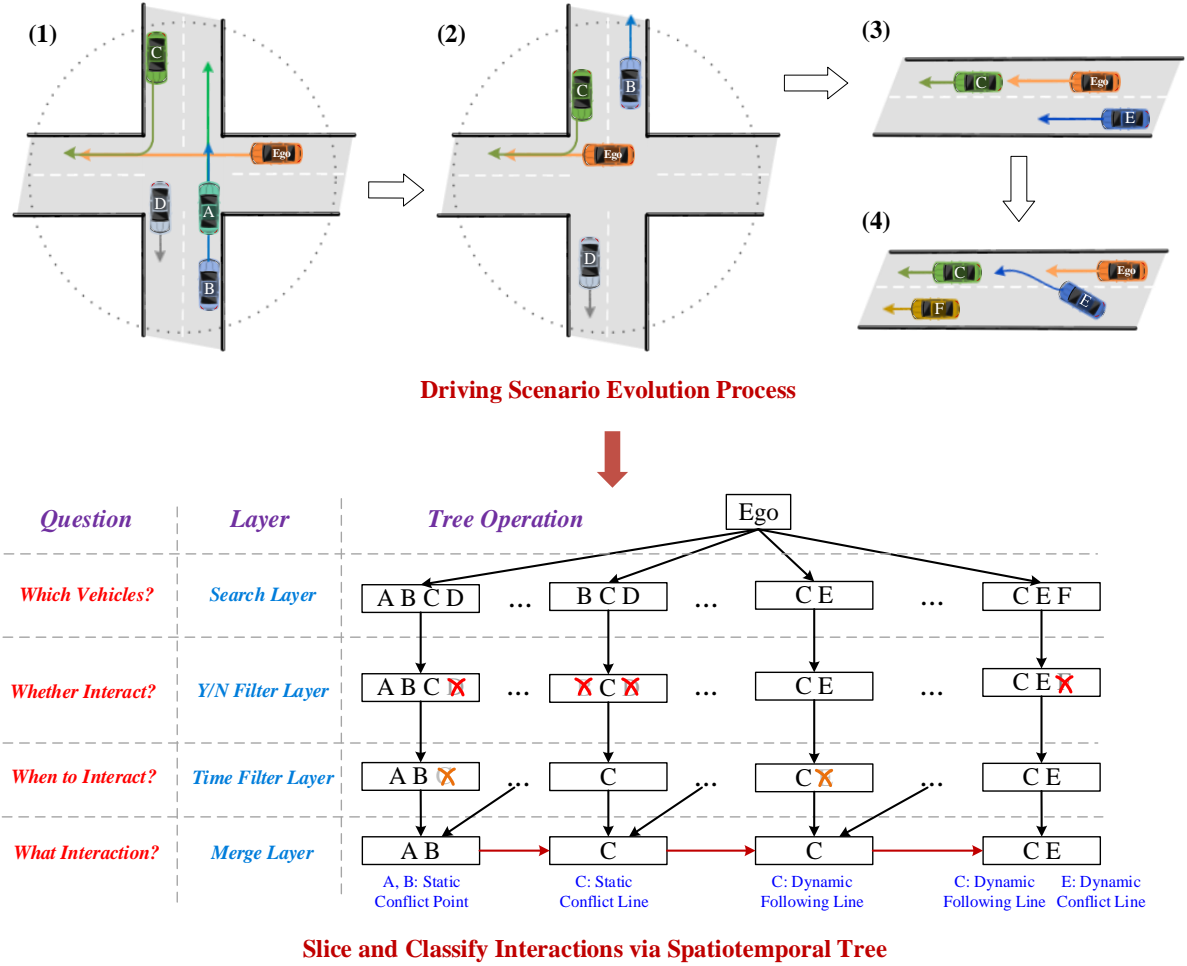

Fig. 3. Interaction-based scenario slicing and classification based on spatiotemporal tree.

*2) Static Conflict Point.* It generally refers to the cross-flow of vehicles in static conflict areas, such as straight and turning flow from different directions at the intersection. During the whole process, the ego interacts with other vehicles that have trajectory crossing points within the conflict area, and determines the pass-orders collaboratively.

Above abstraction of the basic types of vehicle interactions allows us to fuse the road geometry and the interactions caused by vehicle motion flow, which can facilitate further scenario comparison and analysis. All complex interactions can be represented as the combinations of above types.

## IV. INTERACTION BASED SCENARIO SEGMENT AND CLASSIFICATION VIA SPATIOTEMPORAL TREE

We classify driving scenarios based on vehicle interaction types, and propose scenario spatiotemporal tree that slices the scenario data, where each scenario segment contains complete and relatively consistent decision process. We define the scenario segment as the concept of atom scenario *AS* [8], as the following 5-tuple shows:

$$AS_T(Ego, AScene^t, \Psi, B, I)_{t \in T} \qquad (1)$$

where $T$ is the duration of atom scenario, *Ego* is the ego vehicle, $AScene^t$ is the scene of atom scenario at timestamp $t$, which is also called transient atom scenario or a frame of atom scenario, $\Psi$ is the set of driving tasks that the ego needs to complete, $B$ is the behavior of the ego, $I$ is the interaction of the ego with other traffic participants and road environment during the atom scenario.

The atom scenario can be divided into a series of atom scenes $AScene^t$ by time. The temporal evolutions of $AScene^t$ can reflect the driving task execution process. The semantic description for $AScene^t$ should contain the attributes and states of scenario objects, and the relationships between them. The $AScene^t$ is further defined as a 4-tuple:

$$AScene^t(AS) = \left( Ego, E^t(AS), R^t(AS), S^t(AS) \right) \qquad (2)$$

where *Ego* is the ego vehicle, $E^t(AS)$ is the set of surrounding scenario elements (referring to the components of the driving scenario such as vehicles and road network nodes) that interact with the ego vehicle at timestamp $t$, $R^t(AS)$ is the set of semantic relationships between the scenario elements, $S^t(AS)$ is the set of the traffic participants states. We use semantic graph to represent $AScene^t$, and the data structure of atom scenario is a series of graph data. The nodes in the semantic graph represent scenario elements, storing information such as attributes $E^t(AS)$ and states $S^t(AS)$. The edges store the relationships $R^t(AS)$ between the graph nodes.

The more detailed introduction is listed in our previous work [8]. In the previous work, the time parameters and distance thresholds for slicing the atom scenarios are set as relatively fixed values. In this paper, we improve and conduct refined scenario slicing based on vehicle interactions. To simplify the expression, such sliced scenario segments are uniformly referred to atom scenarios.

The detailed slicing operations are shown in Fig. 3. The extraction and classification of vehicle interactions need to solve the "4W" questions, which correspond to the 4 layers of scenario spatiotemporal tree. Each scenario frame, i.e., scene needs to go through the operations of 4 layers.

*1) Search Layer.* At each $AScene^t$, the system extracts the ego's surrounding vehicles within a certain spatial scope, which is determined by the characteristics and distributions of the driving environment. The system stores the vehicles' state data and relation data over a period of time in the buffer. The vehicles have potential interactions with the ego and require further judgment.

*2) Y/N Filter Layer.* Through the extracted data and annotation of lane zones, we detect whether the goals or trajectories of searched vehicles have overlap or intersection with the ego's future flow. If exists, the vehicle will be temporally preserved as interactive vehicle with the ego in the buffer. If not, the vehicle will be removed as non-interactive vehicle.

*3) Time Filter Layer.* The layer determines whether the vehicle at current $AScene^t$ has started or finished interacting with the ego vehicle. We can extract the vehicles' typical states over a period of time horizons, such as lateral velocity or orientation in lane changing process, and steering angle in turning process. According to previous researches [61]-[62], we set a time window and judge whether all the state values in the window meet the threshold requirements. The time window methods can efficiently help determine whether vehicles start or finish executing corresponding actions and interacting with the ego.

To avoid the perturbations, the data should be smoothed and de-noised. To avoid the underestimation of the duration, the threshold should be set as relatively low value to ensure $AS$ contains the complete interaction process. If the vehicle is judged in the interaction process at current scenario frame, it will be stored as interactive vehicle. If not, it will also be temporally removed and wait for judgment at future scenario frames.

*4) Merge Layer.* The layer compares the vehicle interactions in the continuous $AScene$ based on interactions information extracted from above layers, and then merges the scenario frames into a scenario segment $AS$. The condition that the

$AScene^t$ can be merged is that the interaction types between vehicles and the corresponding right-of-way assignment zones of the before and after moments are the same. For example, as shown in Fig. 3, the scenario frames corresponding to the ellipses, where the interaction vehicles and interaction types are the same, can be merged. In contrast, the scenarios corresponding to (2) and (3), where the interaction vehicles are the same, but the interaction types are different, thus cannot be merged. Vehicles are numbered according to the order of appearance. When the number of interactive vehicles reaches a certain threshold (e.g. 10 vehicles), the merging process will be stopped and the decision process will also be transferred to a new section.

---

**Algorithm 1 Scenario segment via spatiotemporal tree**

*Input*: Scenario data stream $S$,
      ego vehicle *Ego*, basic interaction types $\{I\}$

*Output*: Scenario segments $\{AS\}$

1. **For** scenario frame $AScene^t$ in $S$ **do**
    /*** *Search Layer* ***/
2.   Extract vehicle set $V_t$, road node set $N_t$ within a certain spatial scope around *Ego* at timestamp $t$.
3.   Extract the vehicle states, goals and relations in a length of future time horizon.
    /*** *Y/N Filter Layer* ***/
4.   **For** searched vehicle $v$ in vehicle set $V_t$ **do**
5.     Detect whether the goal of $v$ has overlap or intersection with the *Ego* future flow.
6.     **If** detection result is *Yes* **then**
7.       Preserve $v$ as interactive vehicle temporally.
8.       Judge interaction types at $AScene^t$ preliminarily.
9.     **Else**
10.      Remove $v$ from $V_t$.
11.    **End If**
12.  **End For**
    /*** *Time Filter Layer* ***/
13.  Set a time window length $L$ and thresholds.
14.  **For** vehicle $v$ in vehicle set $V_t$ **do**
15.    Select critical states according to interaction types $I_t$.
16.    **If** states in $L$ all meet threshold requirements **then**
17.      Preserve $v$ as interactive vehicle.
18.    **Else**
19.
20.      Remove $v$ temporally and wait for future judgment.
21.    **End If**
22.  **End For**
23. **End For**
    /*** *Merge Layer* ***/
24. **For** scenario frame $AScene^t$ in $S$ **do**
25.   **If** $V_t$ is equal to last $V_t$, and $I_t$ is equal to last $I_t$ **then**
26.     Merge $AScene^t$ to current $AS$ with interaction $I_t$.
27.   **Else**
28.     Store current scenario segment $AS$ to $\{AS\}$.
29.     Begin a new scenario segment.
30.   **End If**
31. **End For**

---

The merged continuous scenario frames constitute $AS$ and correspond to a complete decision-making process. Through the scenario spatiotemporal tree, we complete the scenario segment and preliminary classification based on vehicle interaction types.

The overall algorithm can refer to *Algorithm 1*. Suppose the number of judged scenario frame $AScene^t$ is $M$, and the number of searched vehicles is $N$, the slicing process computation complexity is $O(MN+M)$.

## V. A NEW SCENARIO DISTANCE METRIC FOR SCENARIO CALIBRATION AND LABELING

To further compare atom scenario similarity within each interaction type, we design a new scenario distance metric and calibrate the corner extreme scenarios. The distance metric contains two parts, static atom scene comparison and dynamic atom scenario comparison.

### A. Graph Computation Tree Distance for Scene Comparison

To both consider the graph structure relations and node attributes of different dimensions, we use optimal transport distance [63] of corresponding computation trees [64] to compare the similarity of *AScene*. The stability and generalization of related methods have been verified in [65]. Here we adapt the metric to the comparison and labeling of atom scenario. The tree comparison can handle the vehicle-to-vehicle, vehicle-to-road interactions and different vehicle numbers. The process is shown in Fig. 4.

First, we construct vehicle-to-vehicle (V2V) computation trees $T_{V2V}=\left(E_R,E_V,R_{V2V}\right)$ and vehicle-to-roadnode (V2N) computation tree $T_{V2N}=\left(E_R,E_N,E_V,R_{V2N}\right)$, which correspond to atom scene graph. $E_R$ refers to the root ego vehicle of the tree, $E_V$ refers to the vehicles that have interactions with the root ego vehicle, $E_N$ refers to the road network nodes that have interactions with the root vehicle, $R_{V2V}$ refers to the relationships and connections between vehicles, and $R_{V2N}$ refers to the relationships and connections between vehicles and road network nodes.

From the semantic graph, we can expand different levels of trees. For V2V computation tree of node $u$, we denote $T_{V2V}^1(u)=u$, and $T_{V2V}^l(u)$ as the level $l$ computation tree of node $u$ constructed by $T_{V2V}^{l-1}(u)$ tree connecting the vehicle neighbors of leaf nodes. For the V2N computation tree of node $u$, we also denote $T_{V2N}^1(u)=u$, and $T_{V2N}^l(u)$ as the level $l$ computation tree of node $u$ constructed by $T_{V2N}^{l-1}(u)$ tree connecting the different types of neighbors of leaf nodes. The level value of the extended computation tree should be set appropriately.

The graphs and three-layer computation trees corresponding to scenario frames are shown in Fig. 4. The nodes of V2V computation tree only contain vehicles, and the edges are the relationships between vehicles. In contrast, in V2N computation tree, the nodes at odd levels are all vehicles, and the nodes at even levels are all road network nodes. The edges are interaction relationships between vehicles and road nodes. The comparison results by layers will be summarized to the ego, so we mainly represent the explicit interactions related to ego vehicle at upper level, and represent the implicit interactions of surrounding vehicles at bottom levels. The whole graph corresponds the whole tree $T_{V2V}^L(ego)$ and $T_{V2N}^L(ego)$.

Next, for the comparison of tree structures, we should keep the number of tree nodes as the same. To this end, we introduce the blank tree nodes. For example, node sets on the trees with $u,v$ as root node, i.e. $\Psi(u),\Psi(v)$ respectively, the comparison will be as follows:

$$\begin{pmatrix}\Psi(u)\\\Psi(v)\end{pmatrix}\Rightarrow\begin{pmatrix}\Psi(u)\bigcup\Psi(0)^{\max(|\Psi(v)|-|\Psi(u)|,0)}\\\Psi(v)\bigcup\Psi(0)^{\max(|\Psi(u)|-|\Psi(v)|,0)}\end{pmatrix} \quad (3)$$

where $\Psi(0)$ is a tree which consisting of a node and no edge, $\Psi(0)^n$ refers to the node set of $n$ blank tree nodes.

Then, we compare the scene according to the two types of computation trees based on optimal transport. Suppose the node sets in one layer of the two compared trees are $\Psi(u)=\{u_i\}_{i=1}^n$ and $\Psi(v)=\{v_i\}_{i=1}^n$ with augmented blank nodes except for root node. We solve the optimal transport problem to calculate the distance, i.e.

$$OT_D\begin{pmatrix}\Psi(u)\\\Psi(v)\end{pmatrix}=\min_{\gamma\in R^{n\times n},\,C\in R^{n\times n}}<C,\gamma>/n$$

$$s.t.\ C_{ij}=D(u_i,v_j) \quad (4)$$

$$\gamma 1_n=1,\ \gamma^T 1_n=1$$

where $\gamma$ is the transportation flow satisfying constraints, and $C$ is the distance matrix of nodes in two compared trees, respectively.

Finally, the distance between two computation trees can be calculated recursively. After aligning the scenario elements, the comparison can be conducted from subtrees to roots. The distance matrix will be updated layer by layer as follows:

$$D(T_{V2V}^l(u),T_{V2V}^l(v))=\|f_u-f_v\|+w_l\cdot OT_D\begin{pmatrix}\Psi_{V2V}(u)\\\Psi_{V2V}(v)\end{pmatrix}$$

$$D(T_{V2N}^l(u),T_{V2N}^l(v))=\|f_u-f_v\|+w_l\cdot OT_D\begin{pmatrix}\Psi_{V2N}(u)\\\Psi_{V2N}(v)\end{pmatrix} \quad (5)$$

where $f_u,f_v$ refer to the features of nodes $u,v$ related to the comparison criteria, $w_l$ refers to layer weight, $\Psi_{V2V}(u)$ and $\Psi_{V2N}(v)$ refer to the tree node sets corresponding to different types of computation trees.

The final distance of two *AScene* graphs can be represented as:

$$TD_{V2V}^L(G_{ego_u},G_{ego_v})=OT_D\begin{pmatrix}\Psi_{V2V}(ego_u)\\\Psi_{V2V}(ego_v)\end{pmatrix}$$

$$TD_{V2N}^L(G_{ego_u},G_{ego_v})=OT_D\begin{pmatrix}\Psi_{V2N}(ego_u)\\\Psi_{V2N}(ego_v)\end{pmatrix} \quad (6)$$

$$TD^L(G_{ego_u},G_{ego_v})=\lambda_{V2V}TD_{V2V}^L(G_{ego_u},G_{ego_v})+\lambda_{V2N}TD_{V2N}^L(G_{ego_u},G_{ego_v})$$

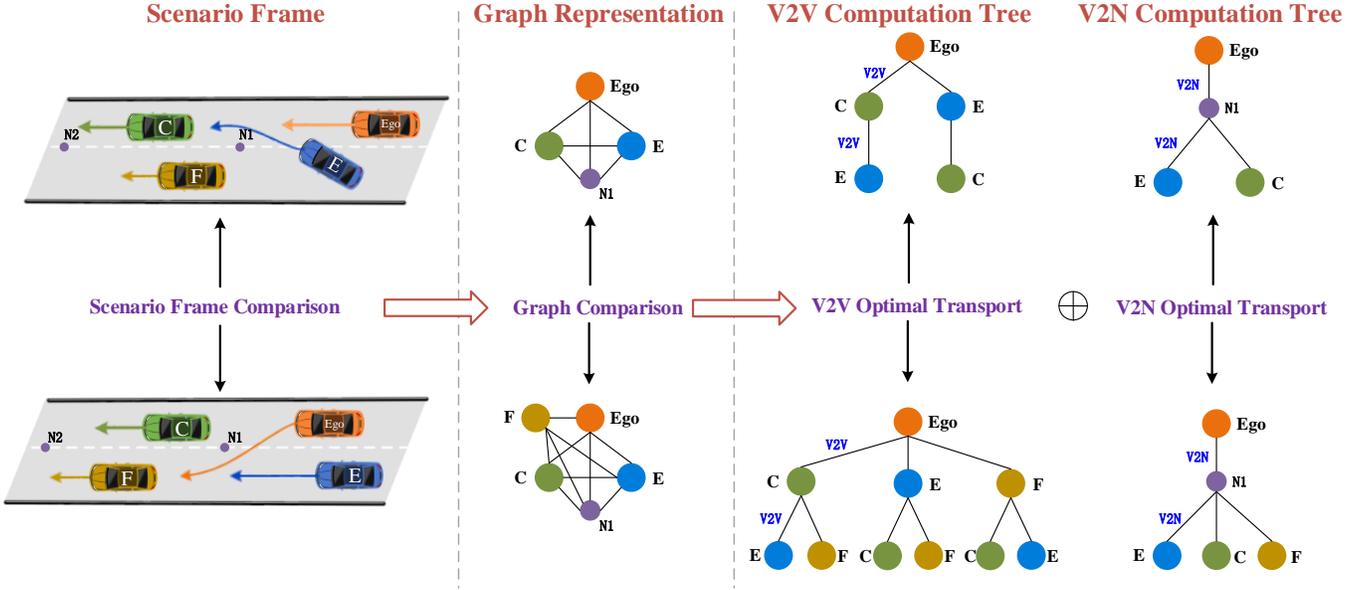

Fig. 4. Scenario frame comparison method based on optimal transport and graph computation tree.

where $G_{ego_A}$, $G_{ego_B}$ are the compared graphs with $u,v$ as ego vehicle, $\lambda_{V2V}$, $\lambda_{V2N}$ are the weights of different types of computation trees, $TD_{V2V}^L$, $TD_{V2N}^L$ are the tree distance of the graph, and $L$ is the level of the constructed tree.

Suppose the maximum number of V2V interaction connections in subtrees is $Y$, the number of comparison times will not exceed $O\left(1+Y^2+\ldots+Y^{2(L-2)}\right)$. Further, to balance the efficiency of comparison and computation, the level $L$ will be set as 3 or 4. The solve of optimal transport will also follow fixed-point network flow implementation [63] to reduce the complexity to $O(Y^2)$. The overall complexity of Computation Tree is $O\left(\left(1+Y^2\right)Y^2\right)$ when $L$ is set as 3. As the non-interactive vehicles have been effectively removed in *Section III*, the interactive vehicle number $Y$ is significantly less than the searched vehicle number $N$, and $Y$ mainly concentrates on 1~4 at each $AScene^t$. Thus the computation complexity will be close to constant.

### B. Graph-DTW for Dynamic Scenario Segments

After calculating the distance of each pair of atom scenes, we further conduct the comparison of dynamic atom scenarios, i.e., scenario segments, which are composed of a series of atom scenes.

The time horizon lengths of sliced scenario segments are usually different. We conduct the comparison of scenario scenarios based on the Dynamic Time Warping (DTW) [30]-[31] algorithm, which corresponds the point of one timestamp of the sequence to the points of multiple consecutive timestamps of the other sequence. By the method of *Section V.A*, we can calculate the distance matrix between each scenario frame of the two graph sequences. Further, as shown in Fig. 5, we find a shortest path in the matrix map from the top left corner to the bottom right corner, and adopt the sum values of elements along the path as the final comparison distance of the two scenarios.

The path update of the DTW method should follow the three constraints: 1) Boundary Condition. The start and end points should be fixed; 2) Continuity. The match process can only be made against adjacent points, not across a point. 3) Monotony. Points along the path proceed monotonically over time and cannot go back to the left. The recurrence rule for the distance is given as follows:

$$L_{\min}(x,y)=D(x,y)+\min\left\{\begin{array}{l}L_{\min}(x-1,y),\\L_{\min}(x,y-1),\\L_{\min}(x-1,y-1)\end{array}\right\} \quad (7)$$

where $D(x,y)$ is the tree distance between $x_{th}$ scenario frame of atom scenario A and $y_{th}$ scenario frame of atom scenario B, $L_{\min}(x,y)$ is the minimum distance updated to the position $(x,y)$ of the distance matrix.

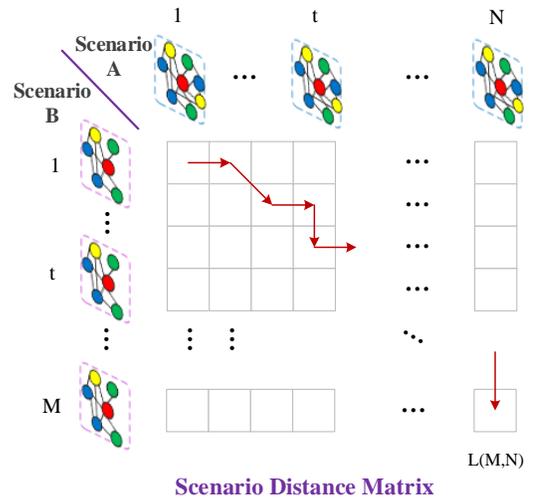

Fig. 5. Dynamic scenario comparison based on dynamic time warping.

Suppose the frame numbers of the two compared graphs are $M$, $N$ respectively. We use the normalized final distance $\dfrac{L_{\min}(M,N)}{\max(M,N)}$ as the similarity degree for atom scenarios comparison. The whole proposed scenario distance metric is referred as Graph-based Dynamic Time Warping (Graph-DTW).

The DTW operation can follow sliding window methods [66] to reduce time and space cost. Suppose the maximum frame number of two scenarios is $M$ and the time window length is set as $W$, the overall Graph-DTW metric computation complexity will be $O\left(W\left(1+Y^2\right)Y^2 M\right) \sim O(CM)$, where $C$ is a constant value.

It is worth mentioning that the framework is relatively flexible. The interaction-based scenario slicing and classifying process is the fixed function, which can help obtain the valid scenario segments. In contrast, the scenario distance metric for refined labeling can be changed appropriately, the selected features for comparison can also be adjusted according to the requirements of concerned criteria. Here we design a feasible metric that properly reflects vehicle interactions, and the effectiveness will be verified by typical cases and testing results on prediction tasks.

## VI. TESTING RESULTS

To verify the effectiveness of the proposed method, we conduct a series of experiments. First, we slice and classify the atom scenarios preliminarily based on vehicle interaction types via spatiotemporal tree. The interaction distribution statistics of typical scenario datasets are conducted. Then, we compare the similarity of atom scenarios base on Graph-DTW method. The corner scenarios can be efficiently labeled compared to other metrics. Finally, we explain the advantages of interaction-based scenario classification via C&C and V&V on typical prediction model. The potential applications for Retrieval Augmented Generation in LLM are also listed.

### A. Datasets and Framework

The experiments require naturalistic driving scenario data, which contain the movement information of traffic participants in a certain spatial range and a continuous time range, as well as the environment information. We adopt the methods on many existing datasets. Here, we select the following typical scenario datasets to demonstrate.

*HighD Dataset.* [67] The dataset contains naturalistic vehicle trajectories recorded on German highways. The data are collected at six different locations with various vehicle maneuvers and different traffic states.

*INTERACTION Dataset.* [68] The dataset contains a variety of rich interaction scenarios, providing vehicle motion data and road map data, etc. The time length of each vehicle's trajectory is relatively long, which is suitable for scenario segment and comparison.

The datasets have been adapted and processed in our scenario framework *MetaScenario* [8]. In the following demonstration,

we focus on the highway scenario in *HighD* Section I, as well as ramp and intersection scenario in *INTERACTION Dataset* to verify the proposed method.

TABLE I
AVERAGE PROCESSING SPEED FOR EACH SCENARIO SEGMENT OF VARIOUS SCENARIOS

| Operation Scenario | Segment via Scenario Tree | Comparison Via Graph-DTW |
|---|---|---|
| *HighD HighWay* | 0.343s | 0.036s |
| *INTERACTION Ramp* | 0.593s | 0.078s |
| *INTERACTION Intersection* | 0.821s | 0.063s |

In the experiments, the machine is equipped with an Intel 10900X CPU with 10 cores. The operating system is Ubuntu 18.04LTS with 64G RAM. In Table I, we test the average processing speed for each scenario segment of various scenarios. The scenario slicing process time for each segment is about 0.3s~0.8s, which will increase a bit with the number of searched vehicles rises. And the comparison of two atom scenarios will cost about 0.03s~0.08s, which is acceptable for later scenario analysis and data indexing.

### B. Scenario and Interaction Distribution Statistics

Based on the scenario spatiotemporal tree, we slice the scenario data stream to several atom scenarios. The atom scenarios are distinguished by different interaction types. We conduct and analyze the scenario and interaction distribution statistics. The results are shown in Fig. 6.

First, from Fig. 6(a) statistics, we count the number of vehicle interactions which exist in each atom scenario. The y-axis is scaled in logarithmic coordinates. Due to the restrictions of vehicle interaction types and relatively sparse traffic flow, the number of interactions in highway environment is relatively small. The number in ramp or intersection scenarios is larger than that of highway scenario. The changes are mainly caused by two reasons: 1) The dense traffic flow. The last two parts of datasets contain the situations of vehicle congestion. 2) The interaction complexity. For interactions in static conflict area, before passing the area, the ego needs to interact with many vehicles that are also about to pass. Both the interaction process and the optimization problem are complex [69].

Second, from Fig. 6(b) statistics, the sliced atom scenarios are with different lengths of durations, which depends on the time of complete decision-making process. Most vehicles can finish the process within 15s. A small number of vehicles will extend the atom scenario to 20s-25s. When the interaction complexity is higher and the interactive vehicle number is larger, the atom scenario also has longer time durations.

Third, from Fig. 6(c) statistics, we count all vehicles in the whole spatiotemporal area, the average number of effectively interactive vehicles, and filtered non-interactive vehicles. The comparison results show that the proposed interaction-based classification method can greatly reduce the scenario space and complexity. Among the various interaction types, the average proportion of filtered non-interactive vehicles exceeds 70%.

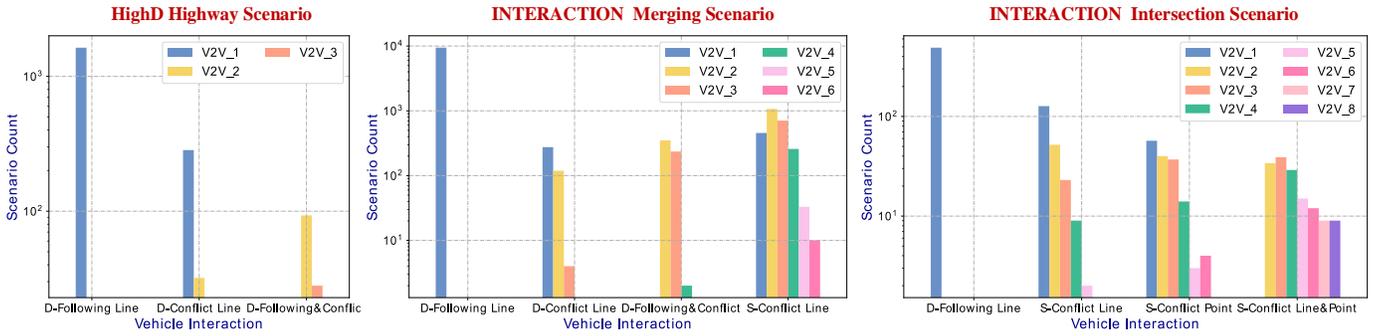

(a). The vehicle interaction statistics of typical scenario datasets. (V2V_n refer to the number of interactive vehicles)

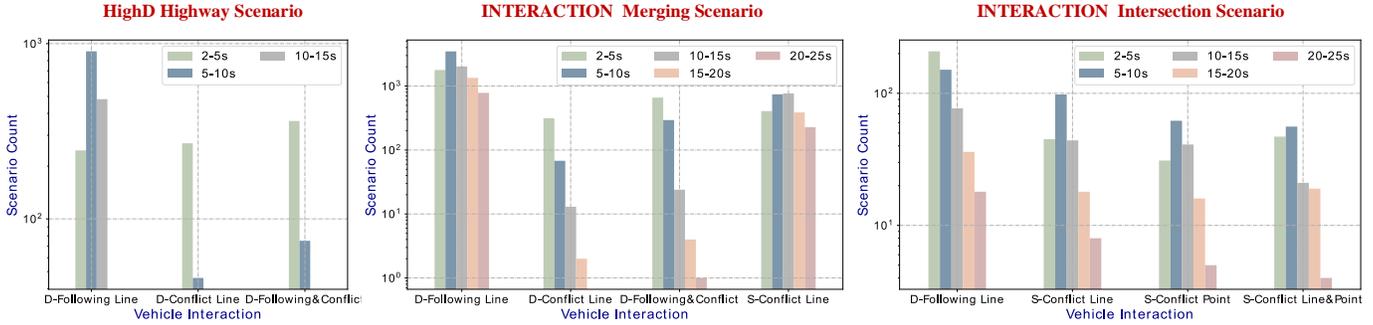

(b). The duration statistics of sliced atom scenarios in typical datasets.

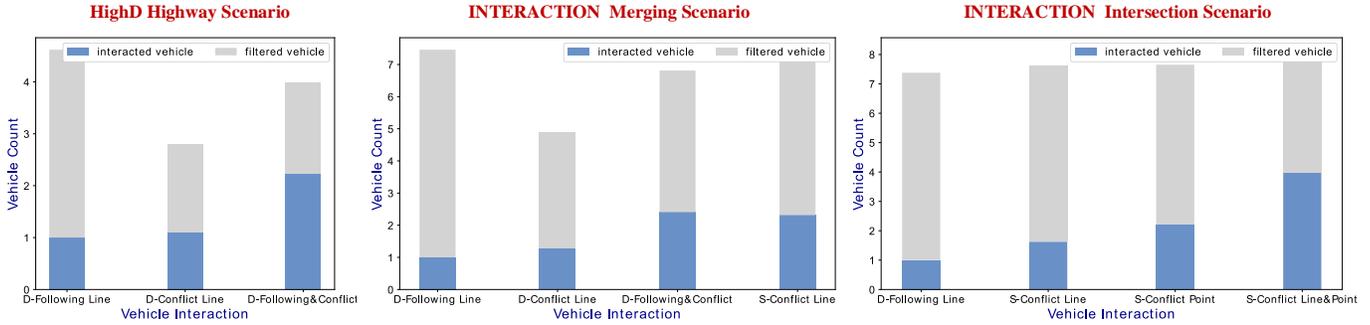

(c). The statistics of effectively interacted vehicles and filtered vehicle in each atom scenario.

Fig. 6. The interaction distribution statistics for sliced atom scenario.

## C. Refined Labeling and Extreme Scenario Extraction

Furthermore, we conduct refined scenario labeling within each interaction type based on proposed Graph-DTW distance metric. The atom scenario distances are calculated without specifying the dimension of scenario representations. Therefore, we use DBSCAN [70], a typical type of density-based clustering methods, to get the dense scenario clusters and filter out the extreme scenarios that are not close to the normal scenario groups. Multi-Dimensional Scaling (MDS) [71] is applied to represent and directly visualize the distance of scenario samples in two-dimensional space.

Driving safety issues are significant aspects that reflect vehicle interactions, here we take driving risk as example to compare scenarios. Generally, driving risk is believed to be caused by suddenly acceleration/deceleration of vehicles, or the obstacles in visual blind area. To specially consider vehicle potential collisions arising from failure to yield the right-of-way [24], which may result from the abnormal movements of vehicles, or disobey of traffic rules.

Since most driving scenarios contain relatively safe driving behaviors, the filtered outliers correspond to the few scenarios with higher risks. The samples are usually at the edge of scenario distribution space [72]-[73], and are more critical and interesting scenarios for researches to take care and conduct testing, which is also verified in our subsequent experiments. Here, we select the level of computation trees as 3 to model the explicit and implicit vehicle interactions with the ego. The tree node feature for scenario comparison related to driving risk is designed as:

$$f = \left( \frac{v}{\Delta r}, \frac{v - v_{ego}}{\Delta r}, \frac{v^2 - v_{ego}^2}{\Delta r} \right) \tag{8}$$

where $v$ is the velocity of the vehicle at each scenario frame, $v_{ego}$ is the velocity of ego vehicle, i.e. the tree root at the scenario frame, $\Delta r$ is the relative distance of the vehicles.

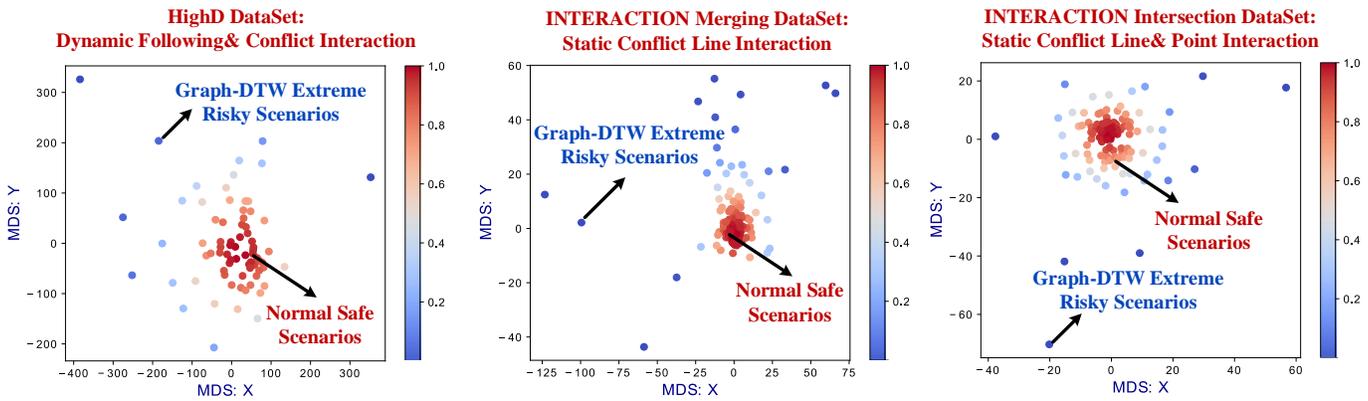

(a). Scenario extraction for each interaction type based on Graph-DTW metric and DBSCAN.

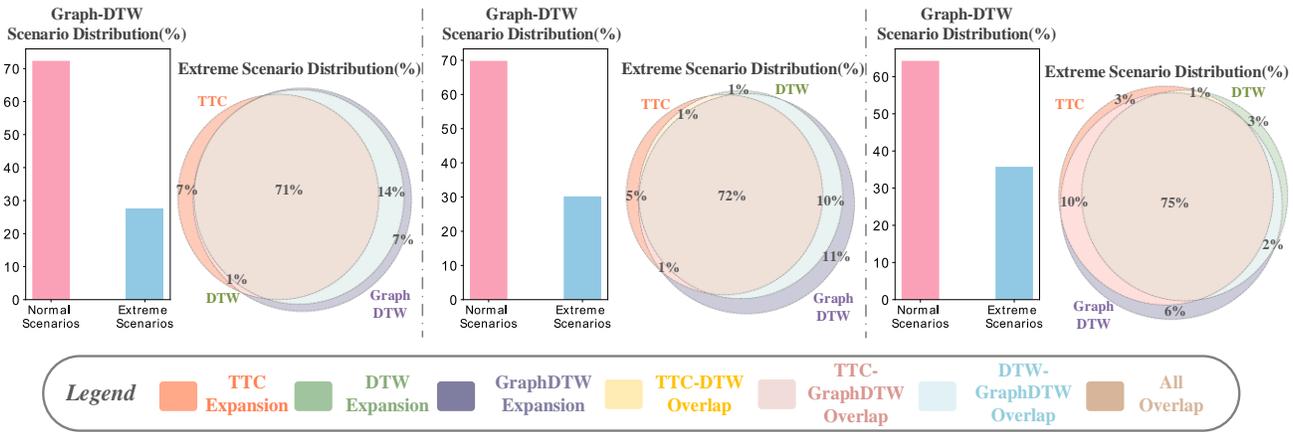

(b). Venn diagrams comparison of extreme scenarios labeled by TTC, DTW, and Graph-DTW metrics.

Fig. 7. Refined scenario labeling under driving risk criteria via Graph-DTW metric.

It is worth mentioning that the relative distance should not be normal Euclidean distance under special interaction area, such as ramps and intersections. Specially, we adopt virtual vehicle mapping method [74]-[75] to set up the virtual leading vehicle along the desired path of ego vehicle when two vehicles are not moving in the same direction to clearly represent the right-of-way assignment.

Among the scenario datasets, we select the most complex interactions to demonstrate the atom scenario comparison and calibration process. As shown in Fig. 7(a), the density of the sample is estimated based on Gaussian kernel function [76] and is displayed as color map with warm color representing dense clusters and cool color representing sparse samples.

To verify the effectiveness, we further compare the extreme risky scenario sets labeled by Graph-DTW metric with the scenario sets labeled by classical TTC (Time-To-Collision) metric and DTW metric. For TTC metric, we set the TTC threshold as 1.0s in highway and ramps [77]-[78], and 0.5s in urban intersections [79] to select the extreme risky scenarios. For TTC metric, we adopt virtual mapping technique to convert the two-dimensional interactions to one-dimensional vehicle-following interactions, then conduct calculation to reflect risks. For DTW metric, we also adopt the aforementioned feature for comparison, zero data arrays will be used to align scenarios that contain different numbers of vehicles and road sampling nodes.

It is notable that the methods here are all under the proposed framework, i.e., the scenario comparison and labeling operations are conducted on the basis of the sliced atom scenarios rather than independently. From Fig. 7, we can observe that:

1) The labeled extreme scenarios occupy small part of the datasets. From Fig. 7(a), we can observe that most of the scenarios are clustered to a center group, and a few scenarios are filtered out due to the large scenario comparison distance. After comparison results with other metrics and manual check, the filtered samples are exactly corner risky scenarios. In listed complex interactions, the proportion of risk scenarios is relatively large with 20%-30%, as shown in Fig. 7(b). In simple interactions, it's even less than 10%.

2) Through the distribution of risky scenarios, we summarize the scenario samples selected by TTC, DTW and Graph-DTW to a full risky set, and analyze the concrete proportion via Venn diagrams. The methods' parameters are selected to obtain the most overlapped scenarios among TTC, DTW and Graph-DTW extractions, and the proportion is finally over 70%. The statistics indicate that the proposed Graph-DTW metric with corresponding designed feature can effectively index the extreme corner scenarios, which correspond to the filtered isolated samples in Fig. 7(a).

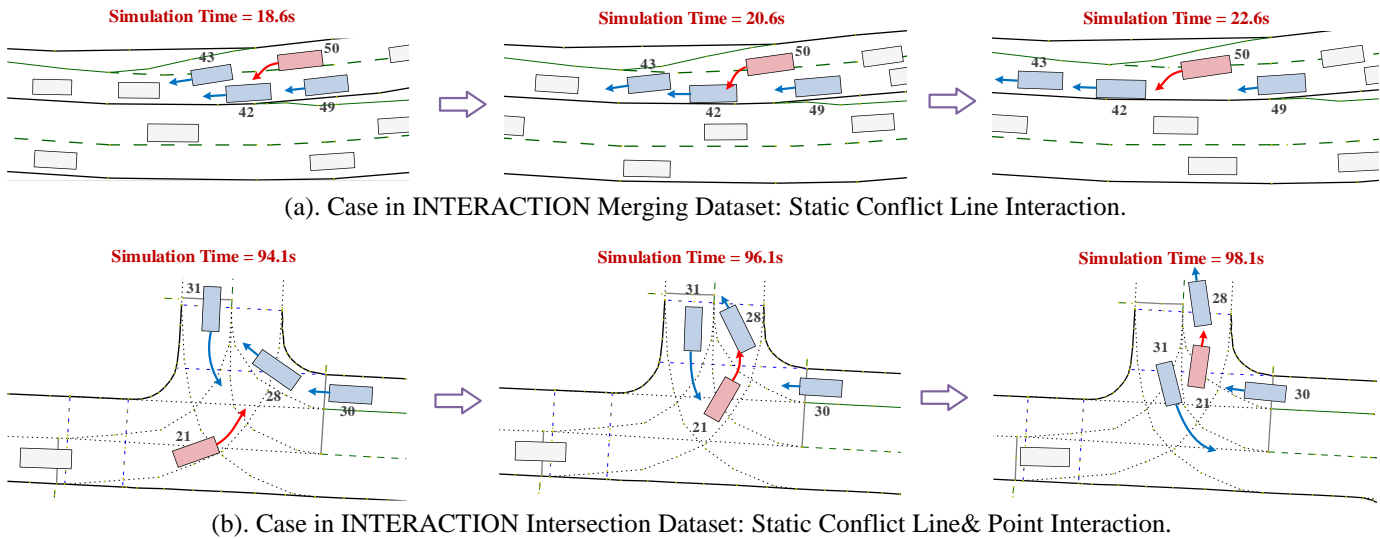

(a). Case in INTERACTION Merging Dataset: Static Conflict Line Interaction.

(b). Case in INTERACTION Intersection Dataset: Static Conflict Line& Point Interaction.

Fig. 8. Typical cases of Graph-DTW expansion risky scenarios.

3) Within the full extreme scenarios, about 6%-11% Graph-DTW expansion scenarios are not detected by TTC and DTW metric. The expansion scenarios mainly contain the risky cases due to violation of proper-right-of-way assignment. As shown in Fig. 8, vehicle 50(a) and vehicle 21(b) both want to seize the right-of-way in the case of extremely tight vehicle spacing in target lanes. In two scenarios, it is obvious that the following vehicles 49(a) and 30(b) are able to get the right-of-way easier. In addition, in Fig. 8(b), vehicle 31(b) is also about to pass the intersection and have conflict with vehicle 21(b). However, during interactions with aggressive vehicles, they all yield the right-of-way to keep safe. Due to the relatively low speed and varying conflict area, TTC metric cannot precisely extract the cases, and due to the insufficient interaction expression, DTW metric cannot extract either. Graph-DTW can characterize interactions among vehicles, and extract the abnormal risky scenarios with complex interactions. The samples can broaden the boundaries of risky scenarios.

### D. Testing Performance Comparisons on Prediction Tasks

In this part, we further provide a typical example for illustrating the usefulness of the labeled samples by the proposed metric and the influences of scenario classification. Based on the classification attributes of scenario samples, we achieve preferential training and testing of extreme scenarios by assigning different weights for different types of samples. The changes of testing loss will be observed to explain the advantages.

Here we take the prediction task as a typical example, and the scenario samples of *INTERACTION* merging dataset are labeled to feed into the model. The historical time horizon and prediction time horizon are both set to 3s, and the time sampling interval are also both set to 0.1s. We totally extract over 20,000 samples for trajectory prediction. The ratio of extreme to normal scenario samples is about 1:5, and the samples of testing set and training set is about 1:4.

The prediction model is adapted from our previous studies [80]-[81], which has been verified to achieve fine performance.

As shown in Fig. 9, Within the attention-based model, $V$ represents the states of ego vehicle and surrounding interactive vehicles, $V_{ego}$ represents the encoder features of ego vehicle, $\hat{V}_0$ is the token to be predicted in the future, and $\hat{V}_{ego}$ is the predicted trajectory of ego vehicle. First, vehicle features are taken as query, key and value, and attention is calculated at the spatial level. The spatial self-attention is operated to be aware of vehicle's spatial relations. In the decoder, we conduct spatial cross-attention to learn refined interactions related to the ego. Then, we extract the temporal interaction by conducting temporal attention. We add a learnable temporal position encoding to fully represent the time series position of vehicles' data. This attention part can help capture the temporal evolutions and make better inferences. Finally, casual convolution layer further predicts the positions along the time axis.

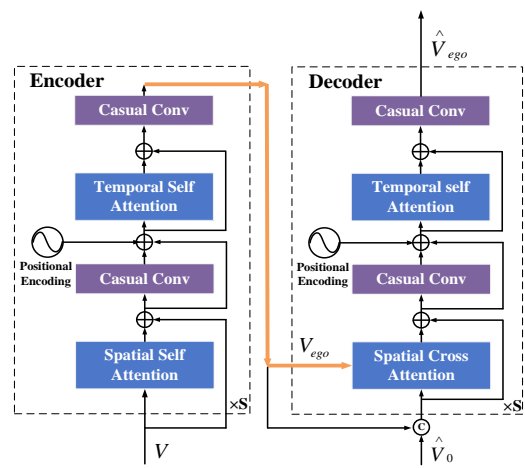

Fig. 9. Attention-based trajectory prediction model.

The prediction performance can be evaluated by the Average Displacement Error (ADE), and Final Displacement Error (FDE) [64]. ADE is the mean Euclidean distance between true positions and predicted positions, and FDE is the Euclidean distance between the final true positions and predicted positions.

The loss function for training is as follows:

$$MAE_i = \frac{1}{F} \sum_{t=1}^{F} \sqrt{\left(p_t^{(x,y)}(i) - \hat{p}_t^{(x,y)}(i)\right)^2}$$

$$MAX_i = \max_{t=1,2...F} \left(\sqrt{\left(p_t^{(x,y)}(i) - \hat{p}_t^{(x,y)}(i)\right)^2}\right) \quad (9)$$

$$Loss = \begin{cases} \lambda_E \left(MAE_i + MAX_i\right) & if \;\; i \in S_E \\ \lambda_N \left(MAE_i + MAX_i\right) & if \;\; i \in S_N \end{cases}$$

where $p_t^{(x,y)}(i)$ is the true vehicle position at future timestamp $t$ of $i_{th}$ scenario samples, $\hat{p}_t^{(x,y)}(i)$ is the corresponding predicted position, $F$ is the prediction time horizon, $S_E$ is the set of extreme scenario samples, $S_N$ is the set of normal scenario samples, and $\lambda_E$, $\lambda_N$ are the corresponding weights for different types of samples. The operations of assigning different weights come from the idea of re-weighting scheme for imbalanced data stream [82].

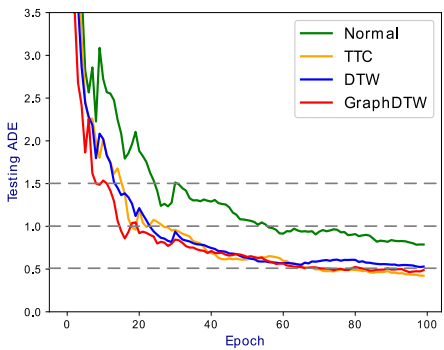

(a). Comparisons of testing ADE for different strategies.

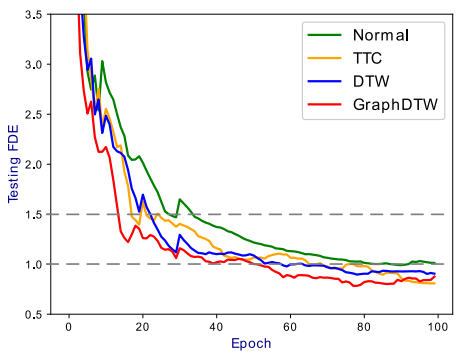

(b). Comparisons of testing FDE for different strategies.

Fig. 10. Testing performance comparisons with training epochs on prediction model.

First, we discuss the advantages of preferentially training and testing the corner scenarios labeled by Graph-DTW. We adopt four sample weight strategies and compare the performance differences. In the normal strategy, we set the weights for all scenarios to the equal. In the TTC, DTW, and Graph-DTW strategies, the weights for their corresponding extracted extreme scenario samples are assigned as a larger value to train and test preferentially. To conduct fair comparison, the weights should be balanced with the number of each sample type. The other parameters, such as batchsize and learning rate, should also be kept the same. The results are averaged through several rounds of experiments.

The changes of testing ADE and FDE for scenarios extracted by different metrics are shown in Fig. 10. Challenging extreme samples provide more valuable information for model training, and larger weights with extreme samples can help resist the noise information from relatively more normal samples in the late training period. With the preferential testing for extreme scenarios, the prediction model converges faster on both ADE and FDE metrics. In addition, the extreme scenario samples labeled by Graph-DTW also achieve faster convergence and better performance compared to others.

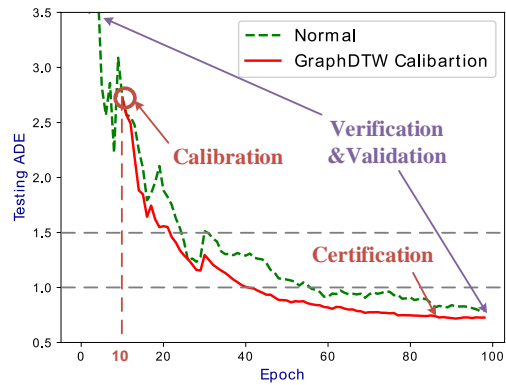

Fig. 11. C&C and V&V for prediction models with Graph-DTW labeled corner scenarios.

Second, we provide the C&C and V&V example for the prediction model. In the initial 10 epochs, we train the model with normal strategies. Then, with the corner scenarios calibrated by Graph-DTW, we preferentially train them and assign larger weights to further calibrate the model with poor convergence. The changes of testing ADE are shown in Fig. 11. It indicates that by highlighting valuable samples, the model parameters can be calibrated for better convergence rate and validation performance. The final well-calibrated and validated model can be certificated as the optimal one for usage in practical applications.

In the above process, C&C and V&V are in couple with each other. On the one hand, by calibrating the model, we can achieve better verification and validation performance. On the other hand, by validating the model, we can determine whether the model is well calibrated and certificate the better one. The above experiments can illustrate the advantages of our proposed metric through prediction tasks, and verify the effectiveness of calibrating and validating AI models with the scenario library constructed by interaction scenario engineering.

### E. Applications for Retrieval Augmented Generation in LLM

With the rise of LLM, many typical autonomous driving tasks, such as prediction, planning, and risk warning, can be automatically and efficiently solved with the strong reasoning and exploring ability of LLM. However, such foundation models still have limitations when dealing with the above knowledge-intensive tasks, which may cause illusion and lag behind task-specific models [83]. Even humans cannot perform well when facing the tasks without relying on external knowledge.

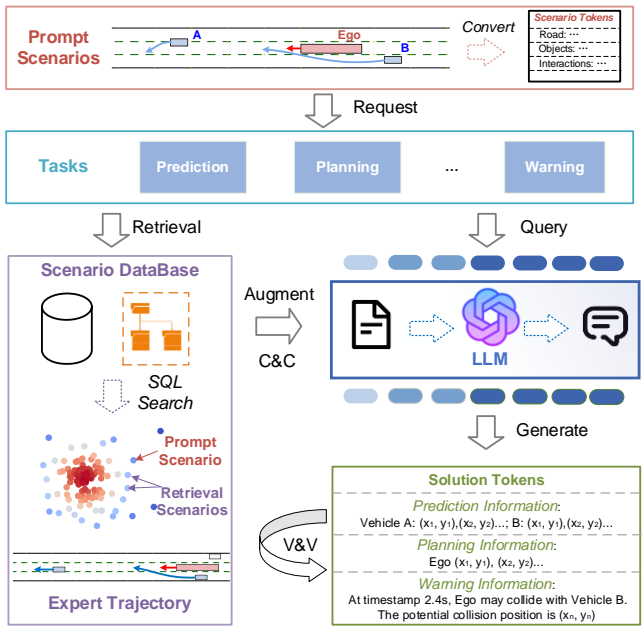

Fig. 12. Applications of *VistaScenario* for RAG.

The scenario library constructed by VistaScenario can help enhance the performance by Retrieval Augmented Generation (RAG), which is shown in Fig. 12.

First, the driving scenarios should be converted to scenario tokens, and request LLM for several tasks. However, the foundation model itself may not generate satisfactory solutions, which needs further calibration and certification.

Next, when requesting the LLM, scenario retrieval is also performed in the constructed scenario database. With SQL search, we can obtain corresponding scenario set that has the same interaction type as the prompt scenario. Moreover, with Graph-DTW scenario distance metric, some scenarios similar to the prompt can be efficiently indexed. Thus, the valuable expert trajectory in these scenarios can help calibrate and augment the LLM. The performance will be enhanced by in-context learning with the retrieval information [84], which provides references and increases the introspective explanation, especially for corner scenario cases.

Finally, LLM generates the solutions for requested tasks. The output tokens will be converted to scenario descriptions and corresponding control signals, and be verified and validated in both simulation and real environments. The *VistaScenario* framework and interactive scenario database can play more important role in the era of foundation models.

## VII. CONCLUSIONS AND FUTURE PERSPECTIVES

In this paper, we propose *VistaScenario* framework to conduct interaction scenario engineering, especially for classification and labeling. We summarize the five basic types of vehicle interactions. The spatiotemporal scenario evolution tree can be effectively applied to slice the scenario stream into series of atom scenarios, which are further classified into different types of interactions. Within each interaction type, we conduct refined comparison and labeling via proposed Graph-DTW metric, and calibrate the extreme atom scenarios and corner cases.

The methods are verified on different types of driving environment and scenario datasets, such as highway, ramps, and intersections. The statistics of each classified interaction type are conducted. Specially, our method can filter out non-interactive vehicles, which occupancy over 70% of scenario space. Through the comparison of Graph-DTW metric and TTC/DTW metric on driving risk criteria, our proposed metric can efficiently extract most of TTC and DTW risky scenarios and expand extreme interaction scenarios by 6%-11%. Testing results on prediction tasks demonstrate the effectiveness of *VistaScenario* framework and the advantages of scenario labeling.

Due to the space limitations, we will discuss the following issues in the future:

First, the interaction scenario engineering framework will be expanded to the land, sea and air integrated transport autonomous systems [85]-[87]. By focusing on the interactions with mixed types of traffic participants, a wide range of valuable scenarios will be indexed and extracted. The *VistaScenario* framework will support the C&C and V&V of next generation transport systems, and promote the collaboration of heterogeneous agents.

Second, the constructed scenario library and corresponding interaction information will be applied for many downstream tasks, such as RAG for LLM, motion planning, risk warning, and intelligent testing [88]-[91] for Vista and transport systems, perhaps in the form of competitions. The advantages of interaction-aware atom scenarios will be further verified in the concrete applications.

## REFERENCES

[1] F.-Y. Wang, P. B. Mirchandani, and Z. Wang, "The VISTA Project and its Applications," *IEEE Intelligent Systems*, vol. 17, no. 6, pp. 72-75, 2002.

[2] F.-Y. Wang, X. Wang, L. Li, and P. Mirchandani, "Creating a Digital-Vehicle Proving Ground," *IEEE Intelligent Systems*, vol. 18, no. 2, pp. 12-15, 2003.

[3] Y. Tian, X. Li, H. Zhang, C. Zhao, B. Li, X. Wang, X. Wang, and F.-Y. Wang, "VistaGPT: Generative Parallel Transformers for Vehicles with Intelligent Systems for Transport Automation," *IEEE Transactions on Intelligent Vehicles*, vol. 8, no. 9, pp. 4198-4207, 2023.

[4] X. Li, P. Ye, J. Li, Z. Liu, L. Cao, and F.-Y. Wang, "From Features Engineering to Scenarios Engineering for Trustworthy AI: I&I, C&C, and V&V," *IEEE Intelligent Systems*, vol. 37, no. 4, pp. 18–26, 2019.

[5] S. Teng, X. Li, Y. Li, L. Li, Z. Xuanyuan, Y. Ai, and L. Chen, "Scenario Engineering for Autonomous Transportation: A New Stage in Open-Pit Mines," *IEEE Transactions on Intelligent Vehicles*, 2024, doi: 10.1109/TIV.2024.3373495.

[6] L. Chen, J. Xie, X. Zhang, J. Deng, S. Ge, and F.-Y. Wang, "Mining 5.0: Concept and Framework for Intelligent Mining Systems in CPSS," *IEEE Transactions on Intelligent Vehicles*, vol. 8, no. 6, pp. 3533-3536, 2023.

[7] J. Li, R. Qin, C. Olaverri-Monreal, R. Prodan, and F.-Y. Wang, "Logistics 5.0: From Intelligent Networks to Sustainable Ecosystems," *IEEE Transactions on Intelligent Vehicles*, vol. 8, no. 7, pp. 3771-3774, 2023.

[8] C. Chang, D. Cao, L. Chen, K. Su, K. Su, Y. Su, F.-Y. Wang, J. Wang, P. Wang, J. Wei, G. Wu, X. Wu, H. Xu, N. Zheng, and L. Li, "MetaScenario: A Framework for Driving Scenario Data Description, Storage and Indexing," *IEEE Transactions on Intelligent Vehicles*, vol. 8, no. 2, pp. 1156-1175, 2023.

[9] L. Chen, Y. Li, C. Huang, Y. Xing, D. Tian, L. Li, Z. Hu, S. Teng, C. Lv, J. Wang, D. Cao, N. Zheng, and F.-Y. Wang, "Milestones in Autonomous Driving and Intelligent Vehicles—Part 1: Control, Computing System Design, Communication, HD Map, Testing, and Human Behaviors," *IEEE Transactions on Systems, Man, and Cybernetics: Systems*, vol. 53, no. 9, pp. 5831-5847, 2023.

[10] X. Li, Y. Tian, P. Ye, H. Duan, and F.-Y. Wang, "A Novel Scenarios Engineering Methodology for Foundation Models in Metaverse," *IEEE Transactions on Systems, Man, and Cybernetics: Systems*, vol. 53, no. 4, pp. 2148-2159, 2023.

[11] J. Wang, J. Wu, and Y. Li, "The Driving Safety Field Based on Driver–Vehicle–Road Interactions," *IEEE Transactions on Intelligent Transportation Systems*, vol. 16, no. 4, pp. 2203-2214, 2015.

[12] Y. Yuan, Q. Wang, and X. T. Yang, "Traffic Flow Modeling with Gradual Physics Regularized Learning," *IEEE Transactions on Intelligent Transportation Systems*, vol. 23, no. 9, pp. 14649-14660, 2022.

[13] Y. Huang, J. Du, Z. Yang, Z. Zhou, L. Zhang, and H. Chen, "A Survey on Trajectory-Prediction Methods for Autonomous Driving," *IEEE Transactions on Intelligent Vehicles*, vol. 7, no. 3, pp. 652-674, 2022.

[14] Y. Hu, W. Zhan, and M. Tomizuka, "Scenario-Transferable Semantic Graph Reasoning for Interaction-Aware Probabilistic Prediction," *IEEE Transactions on Intelligent Transportation Systems*, vol. 23, no. 12, pp. 23212-23230, 2022.

[15] C. Chang, K. Zhang, J. Zhang, S. Li, and L. Li, "Driving Safety Monitoring and Warning for Connected and Automated Vehicles via Edge Computing," in *IEEE International Conference on Intelligent Transportation Systems (ITSC)*, 2022, pp. 3940-3947.

[16] J. Zhang, C. Chang, X. Zeng, and L. Li, "Multi-Agent DRL-Based Lane Change with Right-of-Way Collaboration Awareness," *IEEE Transactions on Intelligent Transportation Systems*, vol. 24, no. 1, pp. 854-869, 2023.

[17] M. Althoff, M. Koschi, and S. Manzinger, "CommonRoad: Composable Benchmarks for Motion Planning on Roads," in *IEEE Intelligent Vehicles Symposium (IV)*, 2017, pp. 719–726.

[18] S. Teng, X. Hu, P. Deng, B. Li, Y. Li, Y. Ai, D. Yang, L. Li, Z. Xuanyuan, F. Zhu, and L. Chen, "Motion Planning for Autonomous Driving: The State of the Art and Future Perspectives," *IEEE Transactions on Intelligent Vehicles*, vol. 8, no. 6, pp. 3692-3711, 2023.

[19] L. Li, X. Wang, K. Wang, Y. Lin, J. Xin, L. Chen, L. Xu, B. Tian, Y. Ai, J. Wang, D. Cao, Y. Liu, C. Wang, N. Zheng, and F.-Y. Wang, "Parallel Testing of Vehicle Intelligence via Virtual-Real Interaction," *Science Robotics*, vol. 4, no. 28, 2019.

[20] J. Ge, H. Xu, J. Zhang, Y. Zhang, D. Yao, and L. Li, "Heterogeneous Driver Modeling and Corner Scenarios Sampling for Automated Vehicles Testing," *Journal of Advanced Transportation*, vol. 2022, pp. 1–14, 2022.

[21] J. Ge, J. Zhang, C. Chang, Y. Zhang, D. Yao, Y. Tian, and L. Li, "Dynamic Testing for Autonomous Vehicles Using Random Quasi Monte Carlo," *IEEE Transactions on Intelligent Vehicles*, 2024, doi: 10.1109/TIV.2024.3358329.

[22] Y. Zhu, Z. Li, F. Wang, and L. Li, "Control Sequences Generation for Testing Vehicle Extreme Operating Conditions Based on Latent Feature Space Sampling," *IEEE Transactions on Intelligent Vehicles*, vol. 8, no. 4, pp. 2712-2722, 2023.

[23] L. Li, C. Zhao, X. Wang, Z. Li, L. Chen, Y. Lv, N.-N. Zheng, and F.-Y. Wang, "Three Principles to Determine the Right-of-Way for AVs: Safe Interaction with Humans," *IEEE Transactions on Intelligent Transportation Systems*, vol. 23, no. 7, pp. 7759-7774, 2022.

[24] S. Feng, X. Yan, H. Sun, Y. Feng, and H. X. Liu, "Intelligent Driving Intelligence Test for Autonomous Vehicles with Naturalistic and Adversarial Environment," *Nature Communications*, vol. 12, no.1, pp. 748, 2021.

[25] S. Feng, H. Sun, X. Yan, H. Zhu, Z. Zou, S. Shen, and H. X. Liu, "Dense Reinforcement Learning for Safety Validation of Autonomous Vehicles," *Nature*, vol. 615, no. 7953, pp. 620–627, 2023.

[26] D. Zhao, H. Lam, H. Peng, S. Bao, D. J. LeBlanc, K. Nobukawa, and C. S. Pan, "Accelerated Evaluation of Automated Vehicles Safety in Lane-Change Scenarios based on Importance Sampling Techniques," *IEEE Transactions on Intelligent Transportation Systems*, vol. 18, no. 3, pp. 595-607, 2016.

[27] L. Li and F. -Y. Wang, "Cooperative Driving at Blind Crossings using Intervehicle Communication," *IEEE Transactions on Vehicular Technology*, vol. 55, no. 6, pp. 1712-1724, 2006.

[28] J. Zhang, S. Li, and L. Li, "Coordinating CAV Swarms at Intersections with a Deep Learning Model," *IEEE Transactions on Intelligent Transportation Systems*, vol. 24, no. 6, pp. 6280-6291, 2023.

[29] S. Kolekar, J. de Winter, and D. Abbink, "Which Parts of the Road Guide Obstacle Avoidance? Quantifying the Driver's Risk Field," *Applied Ergonomics*, vol. 89, 2020, Art. no. 103196.

[30] J. Taylor, X. Zhou, N. M. Rouphail, and R. J. Porter, "Method for Investigating Intradriver Heterogeneity using Vehicle Trajectory Data: A Dynamic Time Warping Approach," *Transportation Research Part B: Methodological*, vol. 73, pp. 59-80, 2015.

[31] W. Wang, A. Ramesh, J. Zhu, J. Li, and D. Zhao, "Clustering of Driving Encounter Scenarios Using Connected Vehicle Trajectories," *IEEE Transactions on Intelligent Vehicles*, vol. 5, no. 3, pp. 485-496, 2020.

[32] E. E. Aksoy, A. Abramov, F. Wörgötter, and B. Dellen, "Categorizing Object-Action Relations from Semantic Scene Graphs," in *IEEE International Conference on Robotics and Automation (ICRA)*, 2010, pp. 398–405.

[33] M. Zipfl, M. Jarosch, and J. M. Zöllner, "Traffic Scene Similarity: A Graph-based Contrastive Learning Approach," in *IEEE Symposium Series on Computational Intelligence (SSCI)*, 2023, pp. 221-227.

[34] M. Dupuis, M. Strobl, and H. Grezlikowski, "OpenDrive 2010 and Beyond-Status and Future of the De Facto Standard for the Description of Road Networks," in *Proceedings of the Driving Simulation Conference Europe*, 2010, pp. 231–242.

[35] F Poggenhans, J. H. Pauls, J. Janosovits, S. Orf, M. Naumann, F. Kuhnt, and M. Mayr, "Lanelet2: A High-Definition Map Framework for the Future of Automated Driving," in *IEEE International Conference on Intelligent Transportation Systems (ITSC)*, 2018, pp. 1672-1679.

[36] T. Kondoh, T. Yamamura, S. Kitazaki, N. Kuge, and E. R. Boer, "Identification of Visual Cues and Quantification of Drivers' Perception of Proximity Risk to the Lead Vehicle in Car-Following Situations," *Journal of Mechanical Systems for Transportation and Logistics*, vol. 1, no. 2, pp. 170-180, 2008.

[37] J. Wang, M. Lu, and Li, K, "Characterization of Longitudinal Driving Behavior by Measurable Parameters," *Transportation Research Record*, vol. 2185, no. 1, pp. 15-23, 2010.

[38] S. Feng, Y. Feng, C. Yu, Y. Zhang, and H. X. Liu, "Testing Scenario Library Generation for Connected and Automated Vehicles, Part I: Methodology," *IEEE Transactions on Intelligent Transportation Systems*, vol. 22, no. 3, pp. 1573-1582, 2021.

[39] J. Zhang, H. Pei, X. J. Ban, and L. Li, "Analysis of Cooperative Driving Strategies at Road Network Level with Macroscopic Fundamental Diagram," *Transportation Research Part C: Emerging Technologies*, vol. 135, pp. 103503, 2022.

[40] K. Kalair, and C. Connaughton, "Anomaly Detection and Classification in Traffic Flow Data from Fluctuations in the Flow–Density Relationship," *Transportation Research Part C: Emerging Technologies*, vol. 127, pp. 103178, 2021.

[41] X. Wang, J. Liu, T. Qiu, C. Mu, C. Chen, and P. Zhou, "A Real-Time Collision Prediction Mechanism with Deep Learning for Intelligent Transportation System," *IEEE Transactions on Vehicular Technology*, vol. 69, no. 9, pp. 9497-9508, 2020.

[42] L. Hartjen, R. Philipp, F. Schuldt, F. Howar, and B. Friedrich, "Classification of Driving Maneuvers in Urban Traffic for Parametrization of Test Scenarios," in *9. Tagung Automatisiertes Fahren. Munich*, Germany: Technical Univ. of Munich, 2019.

[43] L. Hartjen, F. Schuldt, and B. Friedrich, "Semantic Classification of Pedestrian Traffic Scenarios for the Validation of Automated Driving," in *IEEE Intelligent Transportation Systems Conference (ITSC)*, 2019, pp. 3696-3701.

[44] X. Lun and T. Tieniu, "Ontology-based Hierarchical Conceptual Model for Semantic Representation of Events in Dynamic Scenes," in *Proceedings of 2nd Joint IEEE International Workshop on Visual Surveillance and Performance Evaluation of Tracking and Surveillance*, vol. 2005, pp. 57–64, 2005.

[45] H. Beglerovic, T. Schloemicher, S. Metzner, and M. Horn, "Deep Learning Applied to Scenario Classification for Lane-Keep-Assist Systems," *Applied Sciences*, vol. 8, no. 12, pp. 2590, 2018.

[46] R. Yu, H. Ai, and Z. Gao, "Identifying High Risk Driving Scenarios Utilizing a CNN-LSTM Analysis Approach," in *IEEE International Conference on Intelligent Transportation Systems (ITSC)*, 2020, pp. 1-6.

[47] S. Mylavarapu, M. Sandhu, P. Vijayan, K. M. Krishna, B. Ravindran, and A. Namboodiri, "Towards Accurate Vehicle Behaviour Classification with Multi-Relational Graph Convolutional Networks," in *IEEE Intelligent Vehicle Symposium (IV)*, 2020, pp. 321–327.

[48] S. Wang, Y. Zhu, Z. Li, Y. Wang, L. Li, and Z. He, "ChatGPT as Your Vehicle Co-Pilot: An Initial Attempt," *IEEE Transactions on Intelligent Vehicles*, vol. 8, no. 12, pp. 4706-4721, 2023.

[49] X. Li, Q. Miao, L. Li, Y. Hou, Q. Ni, L. Fan, Y. Wang, Y. Tian, and F.-Y. Wang, "Sora for Scenarios Engineering of Intelligent Vehicles: V&V, C&C, and Beyonds," *IEEE Transactions on Intelligent Vehicles*, 2024, doi: 10.1109/TIV.2024.3379989.

[50] F.-Y. Wang, Q. Miao, L. Li, Q. Ni, X. Li, J. Li, L. Fan, Y. Tian, and Q.-L. Han, "When Does Sora Show: The Beginning of TAO to Imaginative Intelligence and Scenarios Engineering," *IEEE/CAA Journal of Automatica Sinica*, vol. 11, no. 4, pp. 809-815, 2024.

[51] R. Qin, F.-Y. Wang, X. Zheng, Q. Ni, J. Li, X. Xue, and B. Hu, "Sora for Computational Social Systems: From Counterfactual Experiments to Artificiofactual Experiments with Parallel Intelligence," *IEEE Transactions on Computational Social Systems*, vol. 11, no. 2, pp. 1531-1550, 2024.

[52] J. Kerber, S. Wagner, K. Groh, D. Notz, T. Kühbeck, D. Watzenig, and A. Knoll, "Clustering of the Scenario Space for the Assessment of Automated Driving," in *IEEE Intelligent Vehicles Symposium (IV)*, 2020, pp. 578-583.

[53] F. Kruber, J. Wurst, and M. Botsch, "An Unsupervised Random Forest Clustering Technique for Automatic Traffic Scenario Categorization," in *IEEE International Conference on Intelligent Transportation Systems (ITSC)*, 2018, pp. 2811-2818.

[54] A. Iosifidis, A. Tefas, and I. Pitas, "Graph Embedded Extreme Learning Machine," *IEEE Transactions on Cybernetics*, vol. 46, no. 1, pp. 311-324, 2016.

[55] M. K. Rahman, M. H. Sujon, and A. Azad, "FusedMM: A Unified SDDMM-SpMM Kernel for Graph Embedding and Graph Neural Networks," in *IEEE International Parallel and Distributed Processing Symposium (IPDPS)*, 2021, pp. 256-266.

[56] Haussler D, "Convolution Kernels on Discrete Structures," *Technical report*, Department of Computer Science, University of California at Santa Cruz, 1999.

[57] S. V. N. Vishwanathan, N. N. Schraudolph, R. Kondor, and K. M. Borgwardt, "Graph Kernels," *Journal of Machine Learning Research*, vol. 11, pp. 1201-1242, 2010.

[58] H. Xu, Y. Zhang, C. G. Cassandras, L. Li, and S. Feng, "A Bi-Level Cooperative Driving Strategy Allowing Lane Changes," *Transportation Research Part C: Emerging Technologies*, vol. 120, pp. 102773, 2020.

[59] J. Zhang, J. Ge, S. Li, S. Li, and L. Li, "A Bi-Level Network-Wide Cooperative Driving Approach Including Deep Reinforcement Learning-Based Routing," *IEEE Transactions on Intelligent Vehicles*, vol. 9, no. 1, pp. 1243-1259, 2024.

[60] H. Yu, C. Chang, S. Li, and L. Li, "CD-DB: A Data Storage Model for Cooperative Driving," *IEEE Transactions on Intelligent Vehicles*, vol. 8, no. 1, pp. 492-501, 2023.

[61] Q. Wang, Z. Li, and L. Li, "Investigation of Discretionary Lane-Change Characteristics using Next-Generation Simulation Data Sets," *Journal of Intelligent Transportation Systems*, vol. 18, no. 3, pp. 246–253, 2014.

[62] T. Toledo and D. Zohar, "Modeling Duration of Lane Changes," *Transportation Research Record*, vol. 1999, no. 1, pp. 71–78, 2007.

[63] H. P. Maretic, M. El Gheche, G. Chierchia, and P. Frossard, "GOT: An Optimal Transport Framework for Graph Comparison," in *Advances in Neural Information Processing Systems (NeurIPS)*, 2019, pp. 13876-13887.

[64] C. Y. Chuang, and S. Jegelka, "Tree Mover's Distance: Bridging Graph Metrics and Stability of Graph Neural Networks," *arXiv preprint arXiv:2210.01906*, 2022.

[65] N. Bonneel, M. Van de Panne, S. Paris, and W. Heidrich, "Displacement Interpolation using Lagrangian Mass Transport," in *SIGGRAPH Asia Conference*, 2011, pp. 1-12.

[66] S. Salvador, and P. Chan, "Toward Accurate Dynamic Time Warping in Linear Time and Space," *Intelligent Data Analysis*, vol. 11, no. 5, pp. 561-580, 2007.

[67] R. Krajewski, J. Bock, L. Kloeker, and L. Eckstein, "The HighD Dataset: A Drone Dataset of Naturalistic Vehicle Trajectories on German Highways for Validation of Highly Automated Driving Systems," in *IEEE Intelligent Transportation Systems Conference (ITSC)*, 2018, pp. 2118–2125.

[68] W. Zhan, L. Sun, D. Wang, H. Shi, A. Clausse, M. Naumann, J. Kummerle, H. Konigshof, C. Stiller, A. De La Fortelle, and M. Tomizuka, "Interaction Dataset: An International, Adversarial and Cooperative Motion Dataset in Interactive Driving Scenarios with Semantic Maps," *arXiv:1910.03088*, 2019.

[69] J. Ding, L. Li, H. Peng, and Y. Zhang, "A Rule-Based Cooperative Merging Strategy for Connected and Automated Vehicles," *IEEE Transactions on Intelligent Transportation Systems*, vol. 21, no. 8, pp. 3436-3446, 2020.

[70] E. Schubert, J. Sander, M. Ester, H. P. Kriegel, and X. Xu, "DBSCAN Revisited, Revisited: Why and How You Should (still) Use DBSCAN," *ACM Transactions on Database Systems*, vol. 42, no. 3, pp. 1-21, 2017.

[71] J. D. Carroll, and P. Arabie, "MultiDimensional Scaling," *Measurement, Judgment and Decision Making*, pp. 179-250, 1998.

[72] L. Li, Y. Lin, Y. Wang, and F.-Y. Wang, "Simulation Driven AI: From Artificial to Actual and Vice Versa," *IEEE Intelligent Systems*, vol. 38, no. 1, pp. 3-8, 2023.

[73] L. Li, N. Zheng, and F.-Y. Wang, "A Theoretical Foundation of Intelligence Testing and Its Application for Intelligent Vehicles," *IEEE Transactions on Intelligent Transportation Systems*, vol. 22, no. 10, pp. 6297-6306, 2021.

[74] A. Uno, T. Sakaguchi, and S. Tsugawa, "A Merging Control Algorithm Based on Inter-Vehicle Communication," in *IEEE/IEEJ/JSAI International Conference on Intelligent Transportation Systems*, 1999, pp. 783-787.

[75] T. Sakaguchi, A. Uno, S. Kato, and S. Tsugawa, "Cooperative Driving of Automated Vehicles with Inter-Vehicle Communications," in *IEEE Intelligent Vehicles Symposium*, 2000, pp. 516-521.

[76] S. Węglarczyk, "Kernel Density Estimation and its Application," in *ITM Web of Conferences*, vol. 23, p. 00037, 2018.

[77] M. Mulder and D. A. Abbink, "Correct and Faulty Driver Support from Shared Haptic Control During Evasive Maneuvers," in *IEEE International Conference on Systems, Man, and Cybernetics*, 2011, pp. 1057-1062.

[78] Q. Wu, Q.-j. Xiang, C. Lu, and J. Lu, "Traffic Safety Evaluation of Highway Intersection with the Use of Conflict Severity Concept," in *International Conference on Intelligent Computation Technology and Automation (ICICTA)*, 2008, pp. 574-578.

[79] U. Shahdah, F. Saccomanno, and B. Persaud, "Application of Traffic Microsimulation for Evaluating Safety Performance of Urban Signalized Intersections," *Transportation Research Part C: Emerging Technologies*, vol. 60, pp. 96-104, 2015.

[80] K. Zhang, C. Chang, W. Zhong, S. Li, Z. Li, and L. Li, "A Systematic Solution of Human Driving Behavior Modeling and Simulation for Automated Vehicle Studies," *IEEE Transactions on Intelligent Transportation Systems*, vol. 23, no. 11, pp. 21944-21958, 2022.

[81] C. Chang, J. Zhang, K. Zhang, W. Zhong, X. Peng, S. Li, and L. Li, "BEV-V2X: Cooperative Birds-Eye-View Fusion and Grid Occupancy Prediction via V2X-Based Data Sharing," *IEEE Transactions on Intelligent Vehicles*, vol. 8, no. 11, pp. 4498-4514, 2023.

[82] X. Peng, F.-Y. Wang, and L. Li, "MixGradient: A Gradient-Based Re-Weighting Scheme with Mixup for Imbalanced Data Streams," *Neural Networks*, vol. 161, pp. 525-534, 2023.

[83] X. Ding, J. Han, H. Xu, X. Liang, W. Zhang, and X. Li, "Holistic Autonomous Driving Understanding by Bird's-Eye-View Injected Multi-Modal Large Models," *arXiv preprint arXiv:2401.00988*, 2024.

[84] X. Dai, C. Guo, Y. Tang, H. Li, Y. Wang, J. Huang, Y. Tian, X. Xia, Y. Lv, and F.-Y. Wang, "VistaRAG: Toward Safe and Trustworthy Autonomous Driving Through Retrieval-Augmented Generation," *IEEE Transactions on Intelligent Vehicles*, 2024, doi: 10.1109/TIV.2024.3396450.

[85] F.-Y. Wang and G. N. Saridis, "A Coordination Theory for Intelligent Machines," *Automatica*, vol. 26, no. 5, pp. 833-844, 1990.

[86] L. Li, J. Song, F.-Y. Wang, W. Niehsen, and N.-N. Zheng, "IVS 05: New Developments and Research Trends for Intelligent Vehicles," *IEEE Intelligent Systems*, vol. 20, no. 4, pp. 10-14, 2005.

[87] J. Guo, L. Chen, L. Li, X. Na, L. Vlacic, and F.-Y. Wang, "Advanced Air Mobility: An Innovation for Future Diversified Transportation and Society," *IEEE Transactions on Intelligent Vehicles*, 2024, doi: 10.1109/TIV.2024.3377464.

[88] C. Chang, J. Zhang, K. Zhang, Y. Zheng, M. Shi, J. Hu, S. Li, and L. Li, "CAV Driving Safety Monitoring and Warning via V2X-Based Edge Computing System," *Frontiers of Engineering Management*, vol. 11, pp. 107–127, 2024.

[89] S. Teng, L. Li, Y. Li, X. Hu, L. Li, Y. Ai, and L. Chen, "FusionPlanner: A Multi-Task Motion Planner for Mining Trucks via Multi-Sensor Fusion," *Mechanical Systems and Signal Processing*, vol. 208, pp. 111051, 2024.

[90] J. Wang, Y. Tian, Y. Wang, J. Yang, X. Wang, S. Wang, and O. Kwan, "A Framework and Operational Procedures for Metaverses-Based Industrial Foundation Models," *IEEE Transactions on Systems, Man, and Cybernetics: Systems*, vol. 53, no. 4, pp. 2037-2046, 2023.

[91] J. Zhang, Z. Li, L. Li, Y. Li, and H. Dong, "A Bi-Level Cooperative Operation Approach for AGV-Based Automated Valet Parking," *Transportation Research Part C: Emerging Technologies*, vol. 128, pp. 103140, 2021.

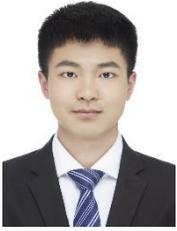

**Cheng Chang** received the B.S. degree from Tsinghua University, Beijing, China, in 2021, where he is currently pursuing the Ph.D. degree with the Department of Automation. His research interests include intelligent transportation systems, intelligent vehicles and machine learning.

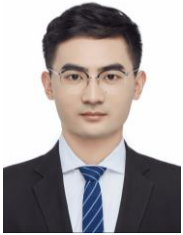

**Jiawei Zhang** received the B.S. degree from Tsinghua University, Beijing, China, in 2020, where he is currently pursuing the Ph.D. degree with the Department of Automation. His current research interests include autonomous driving, intelligent transportation systems, and deep reinforcement learning.

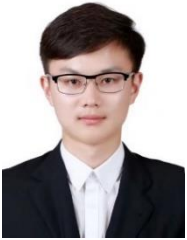

**Jingwei Ge** is currently pursuing the Ph.D. degree with the Department of Automation, Tsinghua University, China. His current research interests focus on intelligent transportation systems, intelligence testing, and autonomous vehicles testing.

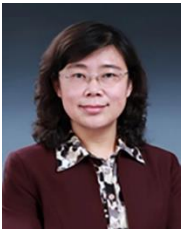

**Zuo Zhang** received the BEng, MEng, and PhD degrees from Tsinghua University, China, in 1989, 1991, and 1995, respectively. She has been a faculty member with Department of Automation, Tsinghua University, after graduation in 1995, with one-year visit to Harvard University from September 1999 to July 2000. Since late 2010, she served as the Deputy Dean of School of Information Science and Technology in Tsinghua University. She is currently a professor with Department of Automation, Tsinghua University. Her current research interests include multi-source data acquisition, fusion and analysis, system control, and optimization techniques for intelligent transportation systems.

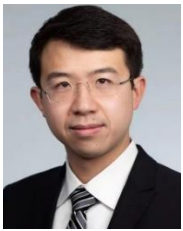

**Junqing Wei** is with DiDi Autonomous Driving, responsible for technical strategy and product R&D. Dr. Wei received B.S degree in automation from Tsinghua Univ. and PhD degree in Electrical and Computer Engineering from Carnegie Mellon University respectively. His research interests include decision making, human-autonomous vehicle interaction under uncertainties and vehicle system architecture. Dr. Wei has been awarded innovator of the year by global automotive tier-1 supplier Aptiv and "best of CES" award for autonomous robotaxi fleet deployment in Las Vegas.

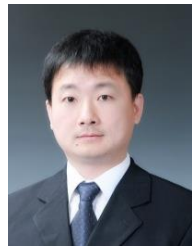

**Li Li** (Fellow, IEEE) is currently a Professor with the Department of Automation, Tsinghua University, Beijing, China, where he was involved in artificial intelligence, intelligent control and sensing, intelligent transportation systems, and intelligent vehicles. He has authored over 170 SCI-indexed international journal articles and over 70 international conference papers. He was a member of the Editorial Advisory Board for *Transportation Research Part C: Emerging Technologies*, and a member of the Editorial Board of *Transport Reviews* and *Acta Automatica Sinica*. He also serves as an Associate Editor for IEEE TRANSACTIONS ON INTELLIGENT TRANSPORTATION SYSTEMS and IEEE TRANSACTIONS ON INTELLIGENT VEHICLES.

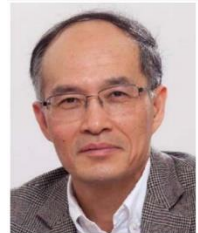

**Fei-Yue Wang** (Fellow, IEEE) received the Ph.D. degree in computer and systems engineering from Rensselaer Polytechnic Institute, Troy, NY, USA, in 1990. He joined The University of Arizona, Tucson, AZ, USA, in 1990 and became a Professor and the Director of the Robotics and Automation Laboratory and the Program in Advanced Research for Complex Systems. In 1999, he founded the Intelligent Control and Systems Engineering Center, Institute of Automation, Chinese Academy of Sciences (CAS), Beijing, China, under the support of the Outstanding Chinese Talents Program from the State Planning Council, and in 2002, was appointed as the Director of the Key Laboratory of Complex Systems and Intelligence Science, CAS. In 2011, he became the State Specially Appointed Expert and the Director of the State Key Laboratory for Management and Control of Complex Systems. His current research focuses on methods and applications for parallel intelligence, social computing, and knowledge automation. Prof. Wang received the National Prize in Natural Sciences of China and became an Outstanding Scientist of ACM for his work in intelligent control and social computing in 2007, the IEEE ITS Outstanding Application and Research Awards in 2009 and 2011, respectively, and the IEEE Norbert Wiener Award in 2014. Since 1997, he has been serving as the General or Program Chair of over 30 IEEE, INFORMS, IFAC, ACM, and ASME conferences. He was the President of the IEEE ITS Society from 2005 to 2007, the Chinese Association for Science and Technology, USA, in 2005, the American Zhu Kezhen Education Foundation from 2007 to 2008, the Vice President of the ACM China Council from 2010 to 2011, and the Vice President and the Secretary General of the Chinese Association of Automation from 2008 to 2018. He was the Founding Editor-in-Chief (EiC) of the International Journal of Intelligent Control and Systems from 1995 to 2000, the IEEE Intelligent Transportation Systems Magazine from 2006 to 2007, the IEEE/CAA JOURNAL OF AUTOMATICA SINICA from 2014 to 2017, and the China's Journal of Command and Control from 2015 to 2020. He was the EiC of the IEEE INTELLIGENT SYSTEMS from 2009 to 2012, the IEEE TRANSACTIONS ON INTELLIGENT TRANSPORTATION SYSTEMS from 2009 to 2016, and the IEEE TRANSACTIONS ON COMPUTATIONAL SOCIAL SYSTEMS from 2017 to 2020, and has been the Founding EiC of Chinese Journal of Intelligent Science and Technology since 2019 and the EiC of the IEEE TRANSACTIONS ON INTELLIGENT VECHICLES since 2022. He is currently the President of CAA's Supervision Council, IEEE Council on RFID, and the Vice President of the IEEE Systems, Man, and Cybernetics Society. He is a Fellow of INCOSE, IFAC, ASME, and AAAS.